\documentclass{aa}  

\usepackage{graphicx}
\usepackage{txfonts}
\newcommand{\cha}{\textit{Chandra}}
\newcommand{\xmm}{XMM-\textit{Newton}}
\newcommand{\nustar}{\textit{NuSTAR}}

\usepackage{amsmath}

\graphicspath{{./}{figures/}}
\begin{document}

   \title{A Simple Method for Predicting $N_H$ Variability in Active Galactic Nuclei}

   %\subtitle{I. Overviewing the $\kappa$-mechanism}

   \author{Isaiah Cox
          \inst{1}
          \and
          Núria Torres-Alb\`a\inst{1}
          \and%\fnmsep\thanks{Just to show the usage
          %of the elements in the author field}
          Stefano Marchesi\inst{1,2}
          \and
          Xiurui Zhao\inst{3}
          \and
          Marco Ajello\inst{1}
          \and
          Andrealuna Pizzetti\inst{1}
          \and
          Ross Silver\inst{1}
          }

   \institute{Department of Physics and Astronomy, Clemson University, 
Clemson, SC, 29634\\
              %\email{wuchterl@amok.ast.univie.ac.at}
         \and
             INAF-Osservatorio Astronomico di Bologna, Via Piero Gobetti, 93/3, I-40129, Bologna, Italy\\
             %\email{c.ptolemy@hipparch.uheaven.space}
             %\thanks{The university of heaven temporarily does not
             %        accept e-mails}
         \and
         Center for Astrophysics | Harvard-Smithsonian, 60 Garden Street, Cambridge, MA 02138, USA
             }

   %\date{Received September 15, 1996; accepted March 16, 1997}

% \abstract{}{}{}{}{} 
% 5 {} token are mandatory
 
  \abstract{The unified model of active galactic nuclei (AGN) includes a toroidal obscuring structure to explain the differences between Type I and Type II AGN as an effect of inclination angle. This toroidal structure is thought to be `clumpy' as the line-of-sight column density, $N_{H}$, has been observed to vary with time in many sources. We present a new method which uses a variation in hardness ratio to predict whether an AGN will have experienced $N_H$ variability across different observations. We define two sets of hard and soft bands that are chosen to be sensitive to the energies most affected by changes in $N_H$. We calculate these ratios for \cha\ and \xmm\ observations on a sample of 12 sources with multiple observations, and compare the predictions of this method with the $N_H$ values obtained from spectral fitting. We find that the method proposed in this work is effective in preselecting sources for variability studies.}

   \maketitle
%
%-------------------------------------------------------------------

\section{Introduction}

   \label{sec:intro}

Active Galactic Nuclei (AGN) are powered by accretion of gas onto supermassive black holes (SMBH) and are among the most luminous sources in the Universe, emitting across the entire electromagnetic spectrum. The unified model for AGN includes an obscuring torus surrounding the accretion disk \citep{antonucci_unified_1993,urry_unified_1995}. Depending on the structure and orientation of the torus, the broad line region (BLR) near the accretion disk may be obscured, resulting in a type II AGN (see e.g. \citealt{hickox_obscured_2018} for a recent review). It was originally thought that this obscuring medium is uniform, however, \cite{krolik_molecular_1988} suggested that this is unlikely. Recent studies of the line-of-sight column density, $N_{H,los}$ (hereafter simply $N_H$), show variability in AGN over timescales ranging from hours \citep[e.g.,][]{elvis_unveiling_2004} to years \citep[e.g.,][]{markowitz_first_2014}. These studies, along with IR SED fitting models \citep[e.g.,][]{2008ApJ...685..160N}, support the idea of a `clumpy' obscuring medium, perhaps made of individual clouds. 

Studying the variability in $N_H$ allows us to constrain properties about the obscuring torus structure such as the density, shape, size, and radial distance of the clouds from the SMBH \citep{risaliti_rapid_2005,maiolino_comets_2010,markowitz_first_2014,pizzetti_multi-epoch_2022,marchesi_compton-thick_2022}. For example, variability on timescales of $\leq1$ day is originated at $\leq10^{-3}$pc (i.e. within the BLR), while monthly and yearly variability likely originates at parsec scales (i.e. in the torus). On the other hand, \cite{laha_variable_2020} looked at a sample of 20 type II AGN and found that 13/20 showed no significant variability in $N_H$ at all, suggesting that the obscuration may be coming from even larger distances associated with the host galaxy. Thus, these studies can provide information about the location of the absorber and the cloud distribution within it. Furthermore, \cite{maiolino_comets_2010} were able to show that the geometry of BLR clouds in NGC 1365 is unlikely to be spherical as is often assumed. 

However, at present, most properties of these clouds remain poorly understood, in large part due to the paucity of sources with known $N_H$ variability available to study. Typically, the way to study $N_H$ variability for AGN with multiple observations is to use some variation of an absorbed powerlaw model to fit the X-ray spectrum \citep[e.g.][]{laha_variable_2020}. Perhaps an even better way is to use a physically-motivated torus model \citep[e.g.,][]{murphy_x-ray_2009, balokovic_new_2018, buchner_x-ray_2019} as done in \cite{pizzetti_multi-epoch_2022} and \cite{marchesi_compton-thick_2022}, for example. However, these methods are time consuming when applied to sources with multiple observations, and are thus not practical for a very large sample of blindly-selected sources. For this reason, very few studies have been performed to date. In fact, the most complete sample of cloud occultation events to date observed only 12 individual events \citep{markowitz_first_2014}, and is still used to calibrate clumpy torus models \citep{buchner_x-ray_2019}.

X-ray data are becoming much more abundant than in the past and could become even more so with future missions such as AXIS \citep{mushotzky_advanced_2019}, Athena \citep{2013arXiv1306.2307N}, and Star-X \citep{2017SPIE10399E..0IS,2022ApOpt..61..505S}. Presently, data is being released from the eROSITA instrument \citep{predehl_erosita_2021} which is expected to detect millions of X-ray point sources, each being observed over timescales ranging from months to years \citep[e.g.][]{salvato_erosita_2022,brunner_erosita_2022}. \cite{marchesi_mock_2020} showed that $90\%$ of the sources detected by AXIS and \textit{Athena} would be first-time detections in the X-rays. Therefore, it is imperative to develop methods to sift through this vast amount of data to pick out observations that are likely to show $N_{H}$ variability. Once these sources are found, they can be studied in depth with the standard spectral modeling techniques. 

A simple measurement that can be used is the hardness ratio (\textit{HR}). The \textit{HR} is a very common measurement that is often interpreted as the X-ray `color' of a source, since it indicates the amount of high-energy (hard) photon counts relative to the low-energy (soft) counts. Because photoelectric absorption is strongly energy dependent, soft X-rays are more likely to be absorbed than hard X-rays. Consequently, large \textit{HR} values typically indicate high $N_H$ values. However, this is not a simple 1:1 relation due to reprocessing effects not related to line-of-sight obscuration.

Previously, hardness ratios have been used on AGN as an indicator of Compton-thickness \citep[e.g.][]{iwasawa_c-goals_2011,torres-alba_c-goals_2018}. Variability in \textit{HR} has also been used to classify AGN \citep[e.g.][]{peretz_classifying_2018} as well as indicate variability in their spectral shape \citep[e.g.][]{2016MNRAS.459.3963C}. However, depending on the choice of `hard' and `soft' bands, it can be difficult to disentangle intrinsic variability in coronal emission and line-of-sight obscuration \citep{caballero-garcia_hard_2012}. By focusing the region of interest on the energies most affected by $N_{H}$ variability, the variability in \textit{HR} is more likely to be due to obscuration effects.

%\textit{HR} variability studies have been performed on AGN samples before as a probe into physical variability. However, due to the choice of `hard' and `soft' bands, they have been unable to disentangle intrinsic variability in the coronal emission and variability in the l.o.s obscuration. By focusing the region of interest on the energies most affected by $N_{H,los}$ variability, the variability in \textit{HR} is more likely to be due to obscuration effects.

In this paper, we present a new method for predicting the variability of $N_H$ between two observations and provide the results as applied to a small sample of carefully analyzed sources. The layout is as follows: In Section \ref{sec:data} we describe the sample of sources and the modeled $N_H$ values used. In Section \ref{sec:analysis}, we describe our method of predicting variation in the modeled $N_H$ values using hardness ratios. In Section \ref{sec:results} we discuss various ways to interpret the reliability of our method and present the results. We summarize our findings in Section \ref{sec:Conclusion}.

\section{Sample and Data} \label{sec:data}

\begin{table*}[t]
  \centering
  \caption{Sample details for \cha\ and \xmm\ data. The best fit $N_H$ values for all three models are in units of $10^{24}$cm$^{-2}$. For details on observations, see TA23.}
  \begin{tabular}{|c|c|ccc|cc|}\hline\hline
    Source Name & Telescope & & $N_H$ & & \multicolumn{2}{c|}{Hardness Ratios}   \\
     & & MYTorus & borus02 & UXCLUMPY & $\textit{HR}_1$ & $\textit{HR}_2$ \\\hline
    3C 452 & \cha & 0.55$^{+0.03}_{-0.03}$ & 0.52$^{+0.02}_{-0.03}$ & 0.44$^{+0.03}_{-0.02}$ & 0.75$\pm0.02$ & 0.18$\pm0.04$  \\
     & \xmm & 0.52$^{+0.03}_{-0.03}$ & 0.49$^{+0.01}_{-0.03}$ &  0.46$^{+0.02}_{-0.02}$ & 0.70$\pm0.01$ & 0.11$\pm0.02$  \\\hline
    3C 105 & \cha & 0.45$^{+0.08}_{-0.05}$ & 0.46$^{+0.04}_{-0.04}$ & 0.49$^{+0.03}_{-0.09}$ & 0.96$\pm0.02$ & 0.18$\pm0.10$  \\
    & \xmm & 0.39$^{+0.05}_{-0.04}$ & 0.39$^{+0.03}_{-0.03}$ & 0.39$^{+0.02}_{-0.03}$ & 0.93$\pm0.05$ & 0.11$\pm0.8$  \\\hline
     NGC 788 & \cha & 0.79$^{+0.08}_{-0.08}$ & 0.73$^{+0.05}_{-0.05}$ & 0.55$^{+0.05}_{-0.02}$ & 0.80$\pm0.03$ & 0.39$\pm0.07$  \\
      & \xmm & 0.82$^{+0.08}_{-0.08}$ & 0.76$^{+0.04}_{-0.04}$ & 0.59$^{+0.08}_{-0.08}$ & 0.82$\pm0.02$ & 0.34$\pm0.03$  \\\hline
     NGC 3281 & \cha & 1.04$^{+0.17}_{-0.17}$ & 0.76$^{+0.10}_{-0.10}$ & 0.76$^{+0.08}_{-0.06}$ & 0.84$\pm0.04$ & 0.53$\pm0.07$  \\
      & \xmm & 1.16$^{+0.17}_{-0.16}$ & 0.86$^{+0.09}_{-0.10}$ & 0.89$^{+0.06}_{-0.07}$ & 0.85$\pm0.02$ & 0.39$\pm0.03$  \\\hline
     IC 4518 A & \xmm\ 1 & 0.21$^{+0.02}_{-0.02}$ & 0.21$^{+0.02}_{-0.01}$ & 0.21$^{+0.08}_{-0.06}$ & 0.73$\pm0.03$ & 0.07$\pm0.05$  \\
      & \xmm\ 2 & 0.31$^{+0.04}_{-0.03}$ & 0.33$^{+0.03}_{-0.03}$ & 0.32$^{+0.01}_{-0.02}$ & 0.70$\pm0.03$ & 0.17$\pm0.04$  \\\hline
     NGC 612 & \cha\ 1 & 1.29$^{+0.29}_{-0.22}$ & 1.27$^{+0.18}_{-0.13}$ & 0.93$^{+0.18}_{-0.19}$ & 0.96$\pm0.06$ & 0.61$\pm0.14$  \\
      & \cha\ 2 & 1.39$^{+0.28}_{-0.22}$ & 1.55$^{+0.19}_{-0.14}$ & 1.10$^{+0.29}_{-0.14}$ & 0.84$\pm0.06$ & 0.71$\pm0.08$  \\
      & \xmm & 0.90$^{+0.11}_{-0.10}$ & 0.89$^{+0.02}_{-0.02}$ & 0.92$^{+0.11}_{-0.13}$ & 0.91$\pm0.04$ & 0.57$\pm0.05$  \\\hline
     NGC 7319 & \cha\ 1 & 0.46$^{+0.04}_{-0.04}$ & 0.47$^{+0.04}_{-0.04}$ & 0.47$^{+0.04}_{-0.05}$ & 0.91$\pm0.02$ & 0.34$\pm0.06$  \\
      & \cha\ 2 & 0.46$^{+0.03}_{-0.03}$ & 0.47$^{+0.03}_{-0.03}$ & 0.46$^{+0.03}_{-0.05}$ & 0.89$\pm0.01$ & 0.28$\pm0.03$  \\
      & \xmm\ 1 & 0.87$^{+0.05}_{-0.05}$ & 0.87$^{+0.06}_{-0.05}$ & 0.84$^{+0.07}_{-0.08}$ & 0.88$\pm0.02$ & 0.36$\pm0.04$  \\\hline
     NGC 4388 & \cha\ 1 & 0.71$^{+0.03}_{-0.03}$ & 0.71$^{+0.04}_{-0.03}$ & 0.66$^{+0.08}_{-0.05}$ & 0.78$\pm0.02$ & 0.38$\pm0.03$  \\
      & \cha\ 2 & 0.91$^{+0.05}_{-0.05}$ & 0.93$^{+0.05}_{-0.04}$ & 0.90$^{+0.04}_{-0.03}$ & 0.73$\pm0.02$ & 0.36$\pm0.03$  \\
      & \xmm\ 1 & 0.37$^{+0.01}_{-0.01}$ & 0.36$^{+0.02}_{-0.01}$ & 0.33$^{+0.01}_{-0.01}$ & 0.81$\pm0.01$ & 0.15$\pm0.02$  \\
      & \xmm\ 2 & 0.235$^{+0.003}_{-0.003}$ & 0.231$^{+0.003}_{-0.003}$ & 0.211$^{+0.002}_{-0.003}$ & 0.72$\pm0.01$ & -0.01$\pm0.01$  \\
      & \xmm\ 3 & 0.267$^{+0.004}_{-0.004}$ & 0.260$^{+0.004}_{-0.004}$ & 0.243$^{+0.003}_{-0.003}$ & 0.78$\pm0.01$ & 0.08$\pm0.01$  \\\hline
     3C 445 & \cha\ 1 & 0.26$^{+0.03}_{-0.01}$ & 0.23$^{+0.01}_{-0.01}$ & 0.22$^{+0.02}_{-0.01}$ & 0.56$\pm0.02$ & 0.12$\pm0.03$  \\
      & \cha\ 2 & 0.33$^{+0.03}_{-0.03}$ & 0.30$^{+0.01}_{-0.01}$ & 0.25$^{+0.02}_{-0.02}$ & 0.53$\pm0.02$ & 0.06$\pm0.04$  \\
      & \cha\ 3 & 0.32$^{+0.03}_{-0.03}$ & 0.28$^{+0.01}_{-0.01}$ & 0.24$^{+0.01}_{-0.01}$ & 0.54$\pm0.02$ & 0.08$\pm0.03$  \\
      & \cha\ 4 & 0.33$^{+0.03}_{-0.03}$ & 0.28$^{+0.01}_{-0.01}$ & 0.25$^{+0.01}_{-0.01}$ & 0.54$\pm0.02$ & 0.14$\pm0.03$  \\
      & \cha\ 5 & 0.31$^{+0.02}_{-0.02}$ & 0.27$^{+0.01}_{-0.01}$ & 0.26$^{+0.01}_{-0.01}$ & 0.54$\pm0.02$ & 0.14$\pm0.03$  \\
      & \xmm & 0.28$^{+0.03}_{-0.03}$ & 0.24$^{+0.01}_{-0.01}$ & 0.20$^{+0.01}_{-0.01}$ & 0.53$\pm0.02$ & 0.00$\pm0.02$  \\\hline
     NGC 835 & \cha\ 1 & 0.89$^{+0.25}_{-0.14}$ & 0.88$^{+0.28}_{-0.14}$ & 1.04$^{+0.18}_{-0.19}$ & 1.00$\pm0.06$ & 0.62$\pm0.24$  \\
      & \cha\ 2 & 0.86$^{+0.32}_{-0.14}$ & 0.85$^{+0.33}_{-0.14}$ & 0.94$^{+0.24}_{-0.16}$ & 1.00$\pm0.06$ & 0.58$\pm0.28$  \\
      & \cha\ 3 & 0.31$^{+0.02}_{-0.03}$ & 0.30$^{+0.03}_{-0.02}$ & 0.28$^{+0.04}_{-0.03}$ & 0.79$\pm0.03$ & 0.29$\pm0.06$  \\
      & \cha\ 4 & 0.32$^{+0.03}_{-0.03}$ & 0.32$^{+0.03}_{-0.03}$ & 0.31$^{+0.04}_{-0.04}$ & 0.81$\pm0.03$ & 0.28$\pm0.08$  \\
      & \cha\ 5 & 0.33$^{+0.03}_{-0.03}$ & 0.32$^{+0.03}_{-0.03}$ & 0.32$^{+0.03}_{-0.03}$ & 0.81$\pm0.02$ & 0.16$\pm0.06$  \\
      & \xmm & 1.53$^{+1.07}_{-0.26}$ & 1.48$^{+1.50}_{-0.23}$ & 1.35$^{+0.05}_{-0.02}$ & 0.62$\pm0.09$ & 0.22$\pm0.13$  \\\hline
     NGC 833 & \cha\ 1 & 0.21$^{+0.07}_{-0.06}$ & 0.19$^{+0.05}_{-0.05}$ & 0.16$^{+0.04}_{-0.03}$ & 0.80$\pm0.06$ & 0.52$\pm0.12$  \\
      & \cha\ 2 & $-$ & $-$ & $-$ & $-$ & $-$  \\
      & \cha\ 3 & 0.33$^{+0.06}_{-0.05}$ & 0.34$^{+0.07}_{-0.06}$ & 0.28$^{+0.05}_{-0.03}$ & 0.77$\pm0.06$ & 0.40$\pm0.13$  \\
      & \cha\ 4 & 0.27$^{+0.05}_{-0.05}$ & 0.27$^{+0.05}_{-0.05}$ & 0.22$^{+0.04}_{-0.04}$ & 0.80$\pm0.07$ & 0.13$\pm0.22$  \\
      & \cha\ 5 & 0.28$^{+0.05}_{-0.04}$ & 0.29$^{+0.05}_{-0.06}$ & 0.24$^{+0.04}_{-0.04}$ & 0.80$\pm0.05$ & 0.26$\pm0.12$  \\
      & \xmm & 0.34$^{+0.07}_{-0.06}$ & 0.31$^{+0.07}_{-0.07}$ & 0.26$^{+0.04}_{-0.03}$ & 0.89$\pm0.10$ & 0.11$\pm0.18$  \\\hline
     4C+29.30 & \cha\ 1 & 0.72$^{+0.16}_{-0.16}$ & 0.68$^{+0.14}_{-0.06}$ & 0.61$^{+0.10}_{-0.11}$ & 0.90$\pm0.07$ & 0.37$\pm0.17$  \\
      & \cha\ 2 & 0.65$^{+0.06}_{-0.06}$ & 0.65$^{+0.06}_{-0.03}$ & 0.61$^{+0.04}_{-0.04}$ & 0.88$\pm0.03$ & 0.22$\pm0.08$  \\
      & \cha\ 3 & 0.59$^{+0.05}_{-0.05}$ & 0.60$^{+0.05}_{-0.01}$ & 0.55$^{+0.04}_{-0.02}$ & 0.88$\pm0.01$ & 0.26$\pm0.04$  \\
      & \cha\ 4 & 0.60$^{+0.06}_{-0.05}$ & 0.60$^{+0.05}_{-0.02}$ & 0.56$^{+0.04}_{-0.02}$ & 0.88$\pm0.02$ & 0.26$\pm0.05$  \\
      & \cha\ 5 & 0.62$^{+0.07}_{-0.06}$ & 0.58$^{+0.05}_{-0.02}$ & 0.54$^{+0.03}_{-0.02}$ & 0.88$\pm0.02$ & 0.29$\pm0.04$  \\
      & \xmm & 0.87$^{+0.18}_{-0.19}$ & 1.08$^{+0.04}_{-0.11}$ & 0.98$^{+0.08}_{-0.10}$ & 0.90$\pm0.04$ & 0.18$\pm0.07$  \\\hline
  \end{tabular}
  \label{tab:sample}
\end{table*}

\begin{table*}[t]
  \centering
  \caption{Sample details for \nustar\ data. The best fit $N_H$ values are in units of $10^{24}$cm$^{-2}$. For details on observations, see TA23.}
  \begin{tabular}{|c|c|ccc|c|}\hline\hline
    Source Name & Telescope & & $N_H$ & & Hardness Ratio  \\
     & & MYTorus & borus02 & UXCLUMPY &  \\\hline
    3C 105 & \nustar\ 1 & 0.45$^{+0.08}_{-0.07}$ & 0.45$^{+0.03}_{-0.03}$ & 0.44$^{+0.03}_{-0.08}$ & 0.41$\pm0.06$ \\
     & \nustar\ 2 & 0.39$^{+0.06}_{-0.06}$ & 0.39$^{+0.06}_{-0.03}$ & 0.40$^{+0.03}_{-0.07}$ & 0.35$\pm0.06$ \\\hline
    NGC 7319 & \nustar\ 1 & 2.17$^{+0.36}_{-0.26}$ & 2.11$^{+0.26}_{-0.22}$ & 0.71$^{+0.25}_{-0.15}$ & 0.38$\pm0.10$ \\
     & \nustar\ 2 & 1.78$^{+0.34}_{-0.34}$ & 1.73$^{+0.30}_{-0.32}$ & 0.98$^{+0.14}_{-0.17}$ & 0.39$\pm0.07$ \\\hline
    NGC 4388 & \nustar\ 1 & 0.30$^{+0.01}_{-0.01}$ & 0.29$^{+0.02}_{-0.02}$ & 0.26$^{+0.02}_{-0.02}$ & 0.21$\pm0.02$ \\
     & \nustar\ 2 & 0.219$^{+0.004}_{-0.005}$ & 0.214$^{+0.004}_{-0.005}$ & 0.195$^{+0.003}_{-0.003}$ & 0.16$\pm0.01$ \\\hline
  \end{tabular}
  \label{tab:sample_nustar}
\end{table*}

\subsection{Sample\label{subsec:sample}}

The sample used to test these methods consists of 12 sources with multiple observations across \cha, \xmm, and \nustar. These sources are studied extensively by Torres-Albà et al. 2023, (hereafter TA23) using the AGN torus models \texttt{borus02} \citep{balokovic_new_2018}, \texttt{MYTorus} \citep{murphy_x-ray_2009}, and \texttt{UXCLUMPY} \citep{buchner_x-ray_2019}, to obtain accurate values of $N_{H}$. The sources are shown in Table \ref{tab:sample} along with the best-fit $N_H$ values found with each of the three models for the \cha\ and \xmm\ observations. Three sources had multiple \nustar\ observations and their information is shown in Table \ref{tab:sample_nustar}. Several sources were found to have observations that vary significantly in $N_{H}$, while others showed no variability. Therefore, this sample has the diversity required to test the predictive power of our hardness ratio method (see Section \ref{subsec:HR}).

\subsection{Data\label{subsec:data}}
\begin{figure}[ht!]
%\plotone{No}
\includegraphics[width=\hsize]{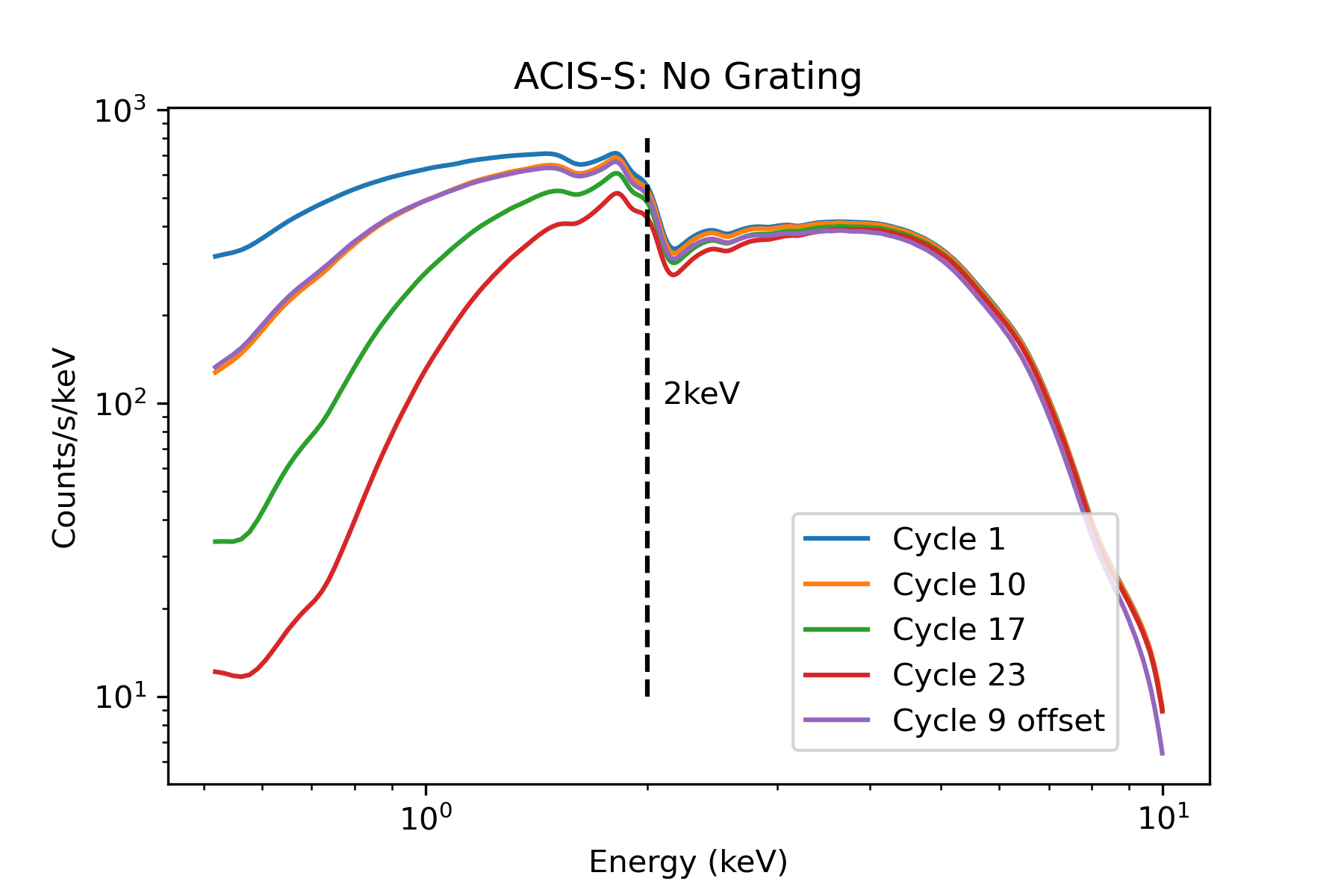}
\centering
\caption{Simulated data of an on-axis source emitting a flat powerlaw ($\Gamma=0$, $\text{norm}=1$) as seen with the ACIS-S camera for cycles 1, 10, 17, and 23. An off-axis source (2.786' from center) is observed in cycle 9 as well.   \label{fig:cycles}}
\end{figure}

This analysis uses observations from \xmm, \cha, and \nustar. For the \xmm\ observations, only the data from the EPIC pn camera \citep{struder_european_2001} is considered due to its higher effective area. All the \cha\ observations were obtained using the ACIS-S camera \citep{garmire_advanced_2003} with no grating.
%\cha\ observations are limited \color{red}(All of the sources in this sample are ACIS-S, but I think we would want to exclude ACIS-I if we did have them because the arf is different.) \color{black} to data obtained using the ACIS-S camera \citep{garmire_advanced_2003} with no grating. 
\cha\ observations range from cycle 1 to cycle 20. However, the degradation in sensitivity with time does not affect our analysis (see Figure \ref{fig:cycles}) because we ignore energies below 2\,keV, where the sensitivity is most significantly reduced.

\cha\ observations may also be affected by vignetting when the source is observed off-axis\footnote{See Figure 6.6 in the \cha\ Proposer's Observatory Guide, \url{https://cxc.harvard.edu/proposer/POG/html/index.html}}. In particular, for sources farther than 5' from the center, there may be a significant softening of the spectrum due to stronger vignetting at higher energies. All of the observations used in this work have the sources of interest within 5' of on-axis. Figure \ref{fig:cycles} also shows simulated data for an off-axis source (2.8' from center) and the relative sensitivity is not significantly reduced until $>$8\,keV where \cha\ is already dominated by background counts. We conclude that the effects from effective area degradation or off-axis sources should not impact this method significantly.

We use data from the FPMA detector for the three sources with multiple \nustar\ observations. We note that there is no substantial difference between the counts observed with FPMA and FPMB, so we choose to consider only FPMA to avoid slightly higher background rates in the FPMB detector.

\section{Method}\label{sec:analysis}

\subsection{Hardness Ratio} \label{subsec:HR}

\begin{figure}[ht!]
%\plotone{No}
\includegraphics[width=\hsize]{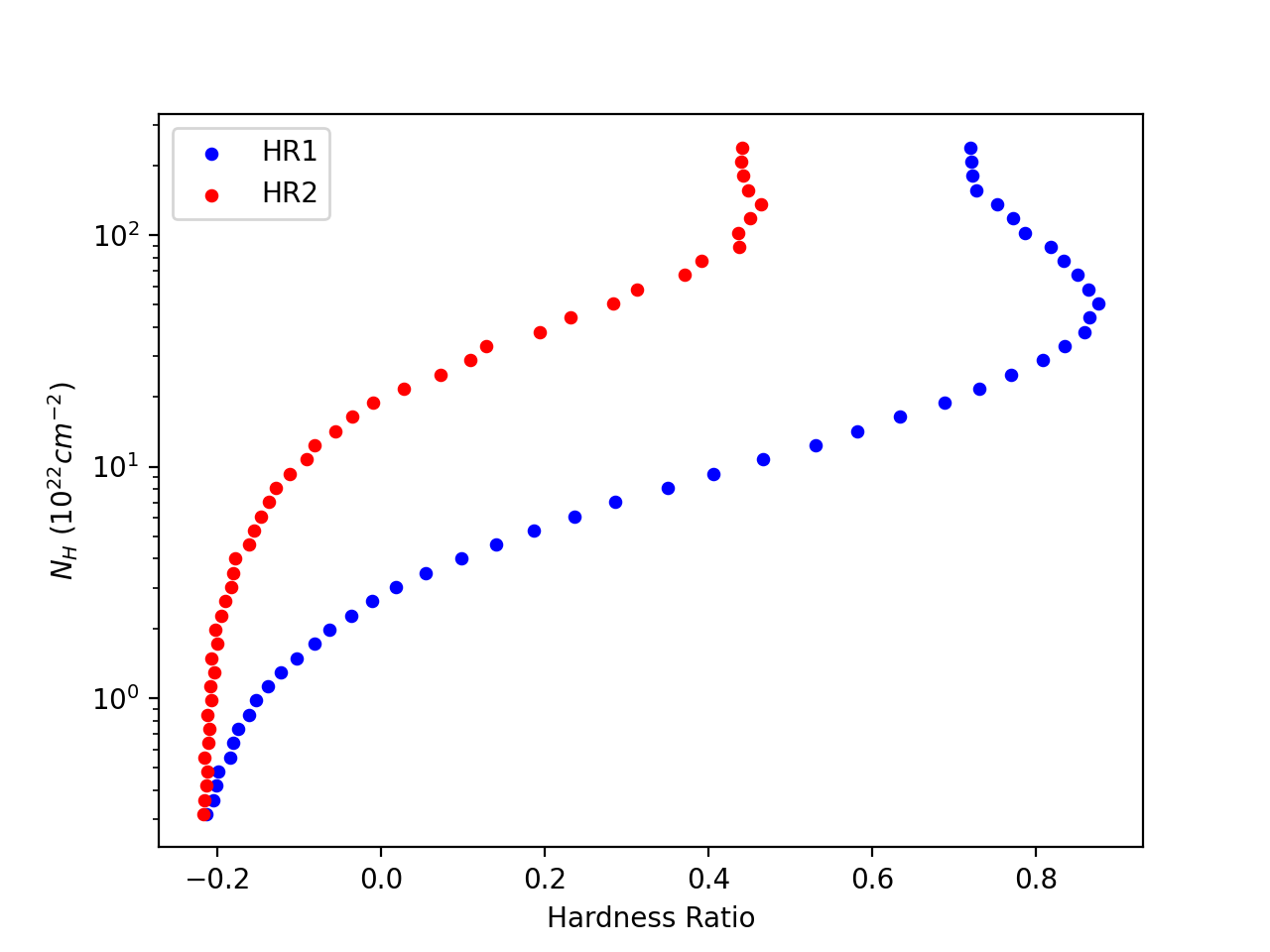}
\centering
\caption{Calculated hardness ratios for data simulated using the \texttt{borus02} model for a range of $N_{H}$ values. $\textit{HR}_2$ continues to increase beyond $N_H\sim3\times10^{23}$cm$^{-2}$ whereas $\textit{HR}_1$ decreases. Neither is sensitive to changes in $N_H$ beyond $\sim10^{24}$cm$^{-2}$ given the selected average torus properties. \label{fig:sim_data_predictions}}
\end{figure}

We define the hardness ratio to be
\begin{equation}
    \textit{HR} = \frac{H-S}{H+S}
\end{equation}
where $H$ and $S$ are the net counts (see eq. \ref{eq:net_bkg}) in the hard and soft bands respectively. We use two different definitions:
\begin{align*}
    \textit{HR}_1: &\text{ Soft  } (\text{2-4\,keV}),\quad \text{ Hard  } (\text{4-10\,keV}) \\
    \textit{HR}_2: &\text{ Soft  } (\text{4-6\,keV}),\quad \text{ Hard  } (\text{6-10\,keV})
\end{align*}
These bands were chosen in an attempt to maximize sensitivity in changes in $N_{H}$. A second hardness ratio, $\textit{HR}_2$, is needed to break a degeneracy present due to the increased importance of the reflection component in sources with high obscuration (see Figure \ref{fig:sim_data_predictions}). Above a certain $N_{H}$, all of the primary soft counts are absorbed, leaving only the reflected counts visible. Since the reflection component does not depend on line-of-sight $N_{H}$, these highly obscured sources show softer $\textit{HR}_1$ as the $N_{H}$ is increased, which decreases the sensitivity of $\textit{HR}_1$ in this $N_{H}$ region and ultimately strips it of its predictive power entirely. According to our simulations using the \texttt{borus02} model, this occurs at $N_{H}\sim3\times10^{23}$cm$^{-2}$ for AGN with photon index, $\Gamma=1.8$; average torus column density, $N_{H,\text{tor}}=10^{24}$\,cm$^{-2}$; and covering factor $c_f=0.5$ (values based on \cite{zhao_properties_2021}). Since $\textit{HR}_2$ is shifted to higher energies, it remains sensitive to $N_{H}$ variability at and beyond this limit as seen in Figure \ref{fig:sim_data_predictions}. It is important to note that these specific quantities should only be taken as indicative since they are meant to represent an `average' AGN, and most individual sources will differ from these simulated data. However, the trends in Figure \ref{fig:sim_data_predictions} should apply for any given source because the average torus properties are not expected to change on the same timescales as line-of-sight $N_{H}$.

The net counts in each band are obtained by setting the data from TA23 in \texttt{XSPEC}. The total counts, $n_{tot}$, along with the fraction of the total count rate that the net count rate contributes, $f$, are recorded. From this information, the net counts ($n_{net}$) and background counts ($n_{bkg}$) are calculated as follows
\begin{equation}\label{eq:net_bkg}
\begin{split}
    n_{net} &= n_{tot}f \\
    n_{bkg} &= n_{tot}(1-f).
\end{split}
\end{equation}

Confidence intervals for \textit{HR} are found by following the methods for Poisson statistics in \cite{gehrels_confidence_1986}. The approximate upper and lower single-sided limits for a measured number of counts $n$ is given by the  \cite{gehrels_confidence_1986} equations (9) and (14)
\begin{equation}\label{eq:limits}
\begin{split}
    n_{u} &\approx (n+1)\left[1-\frac{1}{9(n+1)}+\frac{S}{3\sqrt{n+1}}\right]^3 \\
    n_{l} &\approx n\left(1-\frac{1}{9n}-\frac{S}{3\sqrt{n}}+\beta n^{\gamma}\right)^3.
\end{split}
\end{equation}
where $S=1.645$, $\beta=0.031$, and $\gamma=-2.5$ for $95\%$ single-sided confidence level. This corresponds to a $90\%$ confidence level for a double-sided interval $n_{l}$ to $n_{u}$. These limits are calculated for $n_{tot}$ and $n_{bkg}$ and the error $\delta n$ is taken to be the average difference\footnote{The asymmetry is very small or nonexistent in every case.} between the measured count and the upper and lower bounds
\begin{equation}
    \delta n = \sqrt{\frac{(n_{u}-n)^2+(n-n_{l})^2}{2}}
\end{equation}
The total error on the net counts is then
\begin{equation}
    \delta^2 n_{net} = \delta^2 n_{tot} + \delta^2 n_{bkg}
\end{equation}

This net count error is propagated through the hardness ratio to get the $90\%$ confidence error on \textit{HR} 
\begin{equation}
    \delta^2\textit{HR} = \frac{4}{(H+S)^4}\left[ S^2\delta^2H + H^2\delta^2S \right]
\end{equation}
where $H$ and $S$ are the net counts, $n_{net}$, in the `hard' and `soft' bands respectively.

\subsection{Cross-instrument Comparison\label{subsection:correction}}

The ability to compare observations across multiple instruments is important to maximize the opportunities for variability detection. It is clear that this method should work when comparing \cha\ observations with different \cha\ observations, but it is not as simple when comparing \cha\ with \xmm. In this case, the differences in the instrument response functions make it impossible to meaningfully compare raw hardness ratios between instruments \citep{park_bayesian_2006}. Therefore, a method must be developed to correct for these differences.

\begin{figure}[ht!]
%\plotone{No}
\includegraphics[width=\hsize]{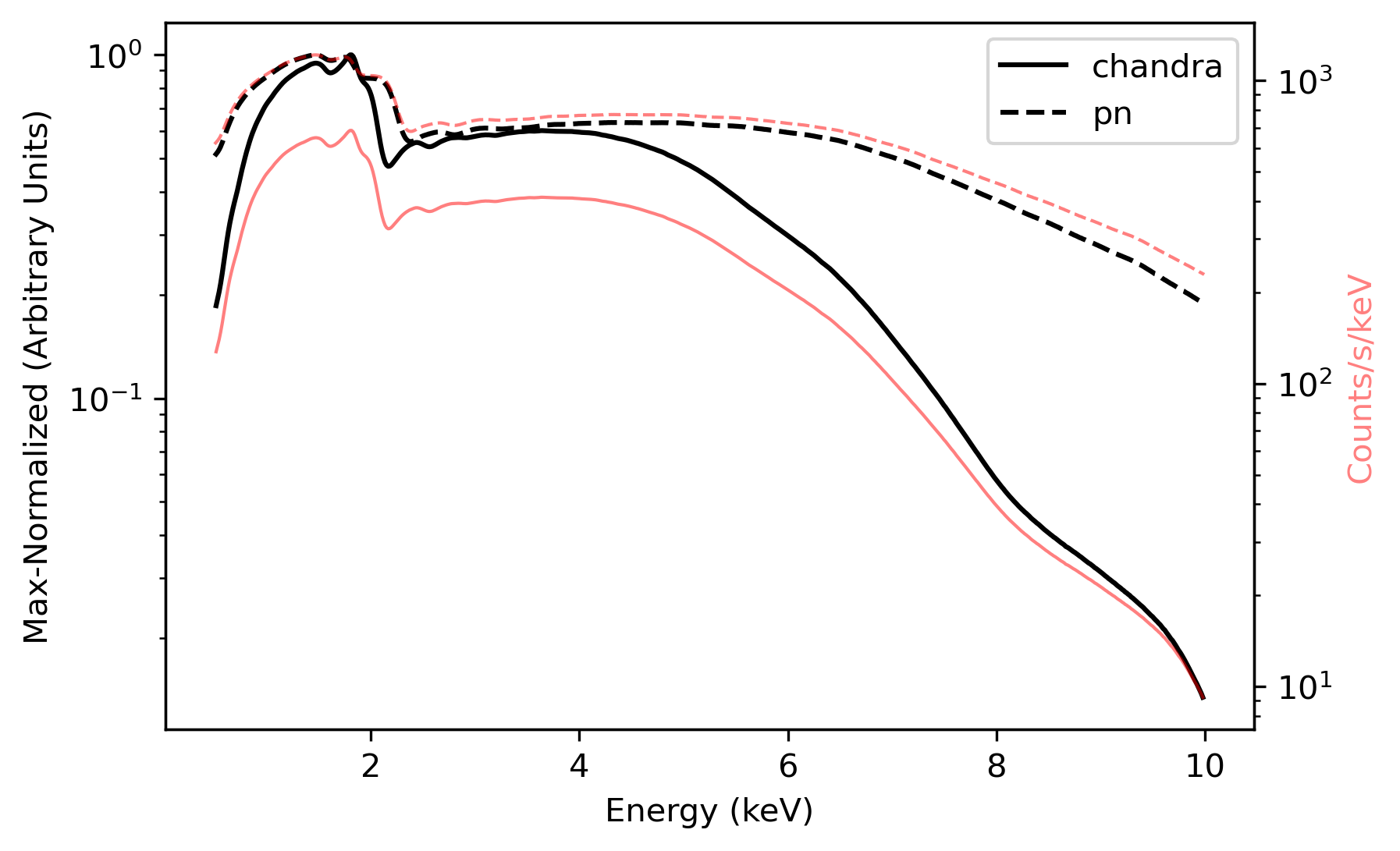}
\centering
\caption{Shape of the response for ACIS-S and pn cameras on an absolute (red) and normalized (black) scale. \cha\ data has a much lower response in the hard band (4-10\,keV) than pn, which needs to be corrected. These data were simulated in the same way as Figure \ref{fig:cycles}. \label{fig:responses}}
\end{figure}

In order to overcome the difficulty in comparing \cha\ to \xmm\ observations, we must account for the differences in the instrument response. In particular, the steep decline of ACIS-S response with respect to EPIC pn response beyond $\sim4$ keV and the complete lack of response in ACIS-S beyond $\sim7$ keV. This is shown in Figure \ref{fig:responses}.

To correct for this difference, we multiply the counts in each band of the \cha\ observations by a correction factor $C_H$ for the hard band and $C_S$ for the soft band. This factor is the ratio of the integrated effective area of the pn detector on \xmm\ to the ACIS-S with no grating on \cha. We find these correction factors to be
\begin{align}
    C_H &= 3.759\quad,\quad (7.185) \nonumber \\
    C_S &= 1.958\quad,\quad (2.368)
\end{align}
for $\textit{HR}_1$, $(\textit{HR}_2)$.

These factors are applied to the count $n$ and count error $\delta n$ after the upper and lower limits have been found. Ultimately, the hardness ratio is \begin{equation}
    \textit{HR} = \frac{C_HH-C_SS}{C_HH+C_SS}
\end{equation}
and the error is
\begin{equation}
    \delta^2\textit{HR} = \frac{4(C_HC_S)^2}{(C_HH+C_SS)^4}\left[ S^2\delta^2H + H^2\delta^2S \right].
\end{equation}

After these corrections, we can more reasonably compare a \cha\ observation to a \xmm\ observation for a given source.

\subsection{\nustar\label{subsection:nustar}}

There are 3 sources in our sample with multiple \nustar\ observations and we applied a modified version of our method to these. The energy bands used to define the hardness ratio for \nustar\ observations are
\begin{align*}
    \textit{HR}_{nu}: &\text{ Soft  } (\text{3-8\,keV}),\quad \text{ Hard  } (\text{8-24\,keV}).
\end{align*}
Since the soft band in this definition covers most of the photons typically absorbed by even highly-obscured AGN \citep[$<10$\,keV;][]{koss_new_2016}, there is no need to introduce a second hardness ratio to break degeneracies.

\subsection{Prediction of $N_{H}$ Variability\label{subsection:prediction}}

It is clear from Figure \ref{fig:sim_data_predictions} that \textit{HR} should depend on $N_{H}$. In this analysis,  we use two different criteria to flag a pair of observations as variable: (1) We take a significant variation in the $90\%$ confidence level of $\textit{HR}_1$ or $\textit{HR}_2$ between the two observations to indicate a significant variation in the $90\%$ confidence level $N_{H}$ for those same two observations. (2) We calculate the $\chi^2$ of each pair of HR values assuming no variability. That is
\begin{equation}
    \chi_{\textit{HR}}^2 = \frac{(\textit{HR}_a-\mu)^2}{\delta^2 \textit{HR}_a} + \frac{(\textit{HR}_b-\mu)^2}{\delta^2 \textit{HR}_b}
\end{equation}
where $\mu$ is the mean $\textit{HR}$ of the two observations $a$ and $b$. The source is flagged as variable if $\chi_{\textit{HR}}^2>2.706$ for either $\textit{HR}_1$ or $\textit{HR}_2$. This value corresponds to a significance level of $\alpha=0.1$. Thereby, we say that the observations are not consistent with each other at the $90\%$ confidence level.

We compare these flagged observations to the `true' variable observations in TA23. Variability in $N_H$ between two observations is defined similarly for Criterion\,1. TA23 obtained the $90\%$ confidence intervals for $N_H$ and these are considered variable if there is no overlap between the values obtained in the two observations. In order to use Criterion\,2, we use the method in \cite{2003sppp.conf..250B} to account for the asymmetry in the $N_H$ errors. Here, we take the $\chi^2$ contribution of an observation to be 
\begin{equation}
\chi^2_i = \frac{N_{H_i}-\mu}{\sigma_i^{+,-}}
\end{equation}
where $N_{H_i}$ is the best fit $N_H$ value for the observation, $\mu$ is the mean best fit $N_H$ value for the two observations, and $\sigma_i^{+}$ and $\sigma_i^{-}$ are the upper and lower limits of the best fit $N_H$. The upper or lower limit is chosen to be in the direction of the mean. For example, if $N_{H_1}<\mu$ then we must have $N_{H_2}>\mu$ and we would calculate
\begin{equation}
    \chi_{N_H}^2 = \left(\frac{N_{H_1}-\mu}{\sigma_1^+}\right)^2 + \left(\frac{N_{H_2}-\mu}{\sigma_2^-}\right)^2.
\end{equation}
As before, we consider two observations to be variable at $90\,\%$ confidence if $\chi_{N_H}^2>2.706$.

%Two $N_H$ values are considered variable if the $90\%$ confidence intervals do not overlap similar to Criterion 1 described above. It is not simple to perform a $\chi^2$ test on these values due to asymmetry in the modeled $N_H$ values so we do not attempt it here. See ----cite Nuria----- for a further exploration of this issue. 

%We compare our variability predictions to the `true' variability in the modeled $N_H$ values from ----cite Nuria----- to determine the reliability of our method. The accuracy of our method is reported in Table \ref{tab:accuracies} for both criteria and all three models.

%Once we have our predictions, we compare them to the model $N_H$ values from -----cite Nuria------. 

\section{Results and Discussion} \label{sec:results}

%\subsection{Individual Sources}\label{subsec:ind}

\subsection{\xmm\ and \cha\ Results}\label{subsec:full}

In total, we had 76 pairs of observations to test our method on and each observation has an $N_H$ value from each of the three models. Since this is a binary classification (variable or not variable), a confusion matrix is one of the best ways to analyze the reliability of the method \citep{stehman1997selecting}. The confusion matrices are shown in Appendix \ref{sec:confusion_matrices}. We consider a true positive, $\textit{TP}$, to be when our method predicts variability \textit{and} the $N_H$ values show variability. A false positive, $\textit{FP}$, is when our method predicts variability, but the $N_H$ values are consistent with each other. A true negative, $\textit{TN}$, and false negative, $\textit{FN}$, are defined similarly.

\subsubsection{Accuracy}\label{subsec:accuracy}
The simplest measure of the reliability would be the accuracy, which is defined as the total number of correct predictions divided by the total number of predictions. In terms of confusion matrix values
\begin{equation}
    \textit{accuracy} = \frac{\textit{TP} + \textit{TN}}{\textit{TP}+\textit{TN}+\textit{FP}+\textit{FN}}.
\end{equation}
The accuracies using each criterion for each of the three models are shown in Table \ref{tab:accuracies} considering only $\textit{HR}_1$, only $\textit{HR}_2$, and both.

The accuracy can be a useful first approximation to the reliability of a method, however, it can hide particular behaviors that are important to note before applying this method to a larger sample. Note that for all three models and for both criteria, using $\textit{HR}_1$ only, seems to make better predictions than considering both ratios especially when using the $N_H$ values derived with the \texttt{MYTorus} model. Looking at the confusion matrices in Appendix \ref{sec:confusion_matrices}, we can see why that is. Only considering $\textit{HR}_1$ is less likely to result in a positive prediction and \texttt{MYTorus} shows more actual negatives (44) than actual positives (32). So, a method that is less sensitive to variability would be expected to do better than a more sensitive method. Considering \texttt{borus02}, which has the same number of actual positives and negatives (38), the accuracy of $\textit{HR}_1$ only and both $\textit{HR}_1$ and $\textit{HR}_2$, are more in agreement. To quantify this, the ``prevalence" is defined as the percentage of actual positives
\begin{equation}
\textit{prevalence} = \frac{\textit{TP}+\textit{FN}}{\textit{TP}+\textit{TN}+\textit{FP}+\textit{FN}}
\end{equation}
The prevalence is also shown in Table \ref{tab:accuracies}. The accuracy of $\textit{HR}_1$ only vs $\textit{HR}_1\&\textit{HR}_2$ seems to be dependent on the prevalence, with lower prevalence favoring $\textit{HR}_1$ only, as expected. In the real-world application of this method, the prevalence will not be known, so it would be hasty to conclude that we should only consider $\textit{HR}_1$ simply because the accuracies are higher for this sample.

%\begin{center}
\begin{table}[t]
\centering
\caption{Accuracies of Criterion\,1 and Criterion\,2 in determining variability, using $N_H$ values from each of the three models. Predictions are based off of variation in $\textit{HR}_1$ alone, $\textit{HR}_2$ alone, and either $\textit{HR}_1$ or $\textit{HR}_2$. Also shown, is the prevalence of each model.}\label{tab:accuracies}

\begin{tabular}{| c| c| c| c| c| }
 \hline
 \multicolumn{2}{|c|}{(Criterion\,1)} & \multicolumn{3}{c|}{Accuracy} \\
 \hline
  Model & \textit{prevalence} & $\textit{HR}_1$ & $\textit{HR}_2$ & $\textit{HR}_1\&\textit{HR}_2$ \\ \hline
 borus & 0.54 & 0.70 & 0.61 & 0.70  \\  
 mytorus & 0.39 & 0.84 & 0.64 & 0.74  \\
 uxclumpy & 0.47 & 0.76 & 0.64 & 0.74 \\
 \hline
 \hline 
 \multicolumn{2}{|c|}{(Criterion\,2)}&\multicolumn{3}{c|}{Accuracy} \\
 \hline
  Model & \textit{prevalence} & $\textit{HR}_1$ & $\textit{HR}_2$ & $\textit{HR}_1 \text{\&} \textit{HR}_2$ \\ \hline
 borus & 0.54 & 0.66 & 0.62 & 0.66  \\  
 mytorus & 0.42 & 0.78 & 0.58 & 0.62  \\
 uxclumpy & 0.50 & 0.70 & 0.61 & 0.64 \\
 \hline
\end{tabular}
\end{table}
%\end{center}

\subsubsection{Precision, Recall, and F-measure}\label{subsec:precision}

Precision and recall can provide a more nuanced interpretation of these results \citep{van1979information}. Precision is a measure of how good the classifier is at avoiding false positives and is defined as
\begin{equation}
    \textit{precision} = \frac{\textit{TP}}{\textit{TP}+\textit{FP}}.
\end{equation}
Recall is a measure of how good the classifier is at finding true positives and is defined as
\begin{equation}\label{eq:recall}
    \textit{recall} = \frac{\textit{TP}}{\textit{TP}+\textit{FN}}.
\end{equation}
Ideally, both of these values would be as close to 1 as possible. However, realistically, this is not achievable and one might want to prioritize one metric over the other. For example, if studying $N_H$ variability in a large sample of sources is the primary goal, precision might be valued over recall to avoid carefully fitting the X-ray spectra of observations that are not variable. On the other hand, if working from a smaller sample, false positives might not be as inconvenient. In this case, one would want to prioritize recall to make sure most of the variable sources are actually flagged. Furthermore, if the hardness ratios are changing, this means that the spectral shape is changing and could indicate something interesting even if it does not happen to be a changing $N_H$. For example, changes in photon index are typically associated to variability of the AGN Eddington ratio, with higher accretion rates corresponding to a softer X-ray spectrum \citep{lu_two_1999,shemmer_hard_2008,risaliti_sdssxmm-newton_2009}.

A single value that accounts for both precision and recall is the $F_{\beta}$-measure \citep{van1979information}. It is defined as
\begin{equation}
    F_{\beta} = (1+\beta^2)\cdot\frac{\textit{precision}\cdot\textit{recall}}{\beta^2\cdot\textit{precision}+\textit{recall}}
\end{equation}
where $\beta>1$ considers recall more important and $\beta<1$ values precision higher. The regular $F$-measure has $\beta=1$ and weights precision and recall equally.

We show the results using $\beta=2$ which values finding truly variable sources over avoiding not variable sources\footnote{The results for $\beta=1$ and $\beta=0.5$ do not provide a very different interpretaion from accuracy.}. The $F_2$ measures are shown in Table \ref{tab:fmeasures}. Here, we see a different interpretation of the results from the standard accuracy shown in Table \ref{tab:accuracies}. In this case, considering variability in either $\textit{HR}_1$ or $\textit{HR}_2$ provides a better score than $\textit{HR}_1$ alone. This is not surprising, as this method is more likely to make a variable prediction and with $\beta=2$, we are artificially rewarding the ability to detect variability.

%\begin{center}
\begin{table}[t]
\centering
\caption{Same as Table \ref{tab:accuracies} but with $F_{2}$-measures instead of accuracies.}\label{tab:fmeasures}

\begin{tabular}{| c| c| c| c| }
 \hline
 \multicolumn{4}{|c|}{\small(Criterion\,1)} \\
 \hline
  Model  & $\textit{HR}_1$ & $\textit{HR}_2$ & $\textit{HR}_1\&\textit{HR}_2$ \\ \hline
 borus & 0.52 & 0.54 & 0.69  \\  
 mytorus & 0.68 & 0.56 & 0.76  \\
 uxclumpy & 0.58 & 0.57 & 0.74 \\
 \hline
 \hline 
 \multicolumn{4}{|c|}{(Criterion\,2)} \\
 \hline
  Model & $\textit{HR}_1$ & $\textit{HR}_2$ & $\textit{HR}_1 \text{\&} \textit{HR}_2$ \\ \hline
 borus & 0.51 & 0.69 & 0.77  \\  
 mytorus & 0.63 & 0.66 & 0.75  \\
 uxclumpy & 0.54 & 0.68 & 0.76 \\
 \hline
\end{tabular}
\end{table}
%\end{center}

\subsubsection{Receiver Operating Characteristic}\label{subsec:roc}

Similar to precision and recall, one can define the false positive rate, \textit{FPR}, and true positive rate, \textit{TPR}. The \textit{FPR} is the ratio of false positives to the total number of actual negatives and is defined as
\begin{equation}
    \textit{FPR} = \frac{\textit{FP}}{\textit{FP}+\textit{TN}}.
\end{equation}
The \textit{TPR} is the ratio of true positives to total actual positives and is equivalent to \textit{recall} (Eq. \ref{eq:recall}). The receiver operating characteristic (ROC) plots the \textit{TPR} against the \textit{FPR} and therefore again provides a measure of how sensitive we are to true positives and how resistant we are against false positives \citep{fawcett2006introduction}. A perfect classifier would be at the point (0,1) while a random classifier would be along the line $\textit{TPR}=\textit{FPR}$. An ROC curve can be obtained by varying the decision threshold.

\begin{figure}[h!]
%\plotone{No}
%\vspace{1cm}
\includegraphics[width=\hsize]{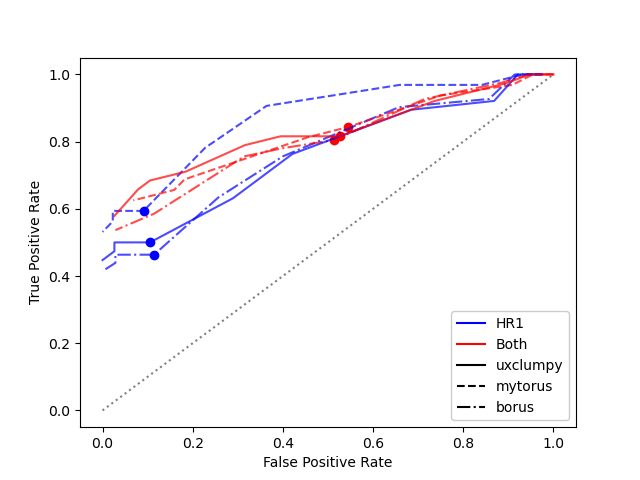}
\centering
\caption{ROC curve for all three models considering both $\textit{HR}_1\&\textit{HR}_2$ and $\textit{HR}_1$ alone. The blue and red dots correspond to the values at 90\,\% confidence level for each combination. The grey dotted line represents a theoretical classifier with no predictive value. \label{fig:roc}}
\end{figure}

Figure \ref{fig:roc} shows the ROC curves for each of the three models considering only variability in $\textit{HR}_1$, as well as variability in both $\textit{HR}_1$ and $\textit{HR}_2$. The $\chi^2$ critical value for one degree of freedom is the decision threshold that was varied to obtain the curves\footnote{This is only changed for the $\textit{HR}$ variability flag. The $N_H$ variability is fixed at $\chi^2=2.706$ or 90\% confidence.}. The values used are $\chi^2=$[0, 0.001, 0.004, 0.016, 0.102, 0.455, 1.32, 2.706, 3.841, 5.024, 6.635, 10.828, 15.137, 19.511] which correspond to confidence levels of $\text{CL}=$[$<$1\,\%, $<$1\,\%, 5\,\%, 10\,\%, 25\,\%, 50\,\%, 75\,\%, 90\,\%, 95\,\%, 99\,\%, 99.9\,\%, 99.99\,\%, 99.999\,\%]. The blue and red points in Figure \ref{fig:roc} show the ROC values at 90\,\% for each classifier. %Simply using the 90\,\% confidence level to match the confidence intervals on the $N_H$ values is acceptable as a predictor, but does   %, while the stars show the best points along the two best classifiers, namely, MYTorus considering variation in only $\textit{HR}_1$ and UXCLUMPY considering variation in both. The confidence levels at these points are 75\,\% and 99\,\%, respectively. 

Here we see that for all combinations the predictive value is much better than random guessing. Furthermore, for confidence levels higher than 90\%, using both ratios leads to better results except for the \texttt{MYTorus} model. This can again be explained with the fact that the \texttt{MYTorus} model has a lower prevalence than \texttt{borus02} or \texttt{UXCLUMPY}, so the method that is more likely to predict no variability will appear better. However, we reiterate that in general, the prevalence will not be known, so using both ratios is likely to be a more robust classifier.

Notably, as the critical value is increased up to $\chi^2>10.282$ corresponding to a confidence level of 99.9\,\%, the \textit{FPR} goes to zero while the \textit{TPR} remains around or above 0.5. This means that by decreasing our sensitivity to variability (up to a point), we can reduce the number of false positives to almost zero, and still be able to detect more than half of the true positives. %See Appendix \ref{sec:confusion_matrices} for the confusion matrices.

\subsection{\nustar\ Results}

\begin{figure*}
    \centering
    \includegraphics[width=0.33\textwidth]{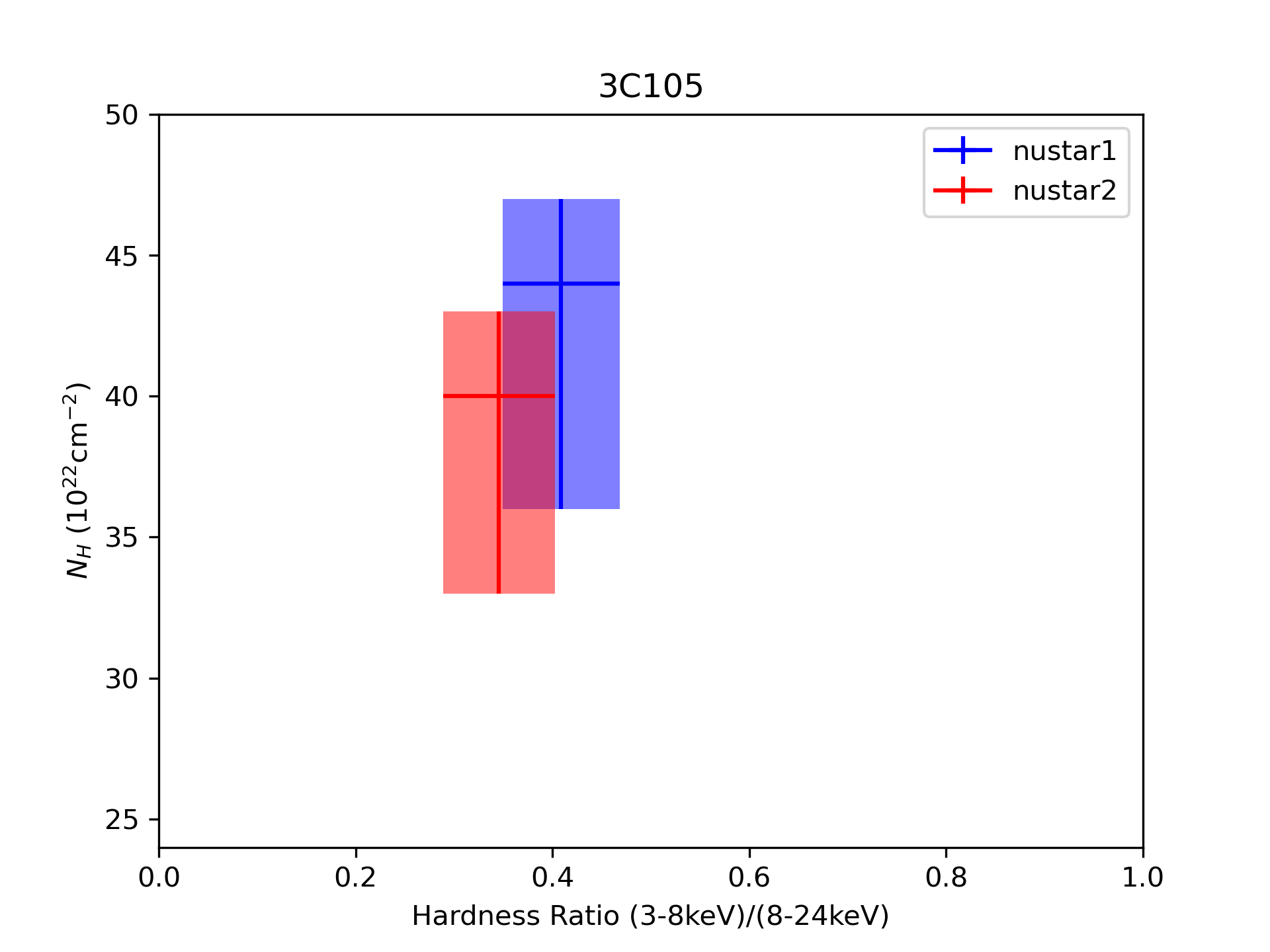}\hfill
    \includegraphics[width=0.33\textwidth]{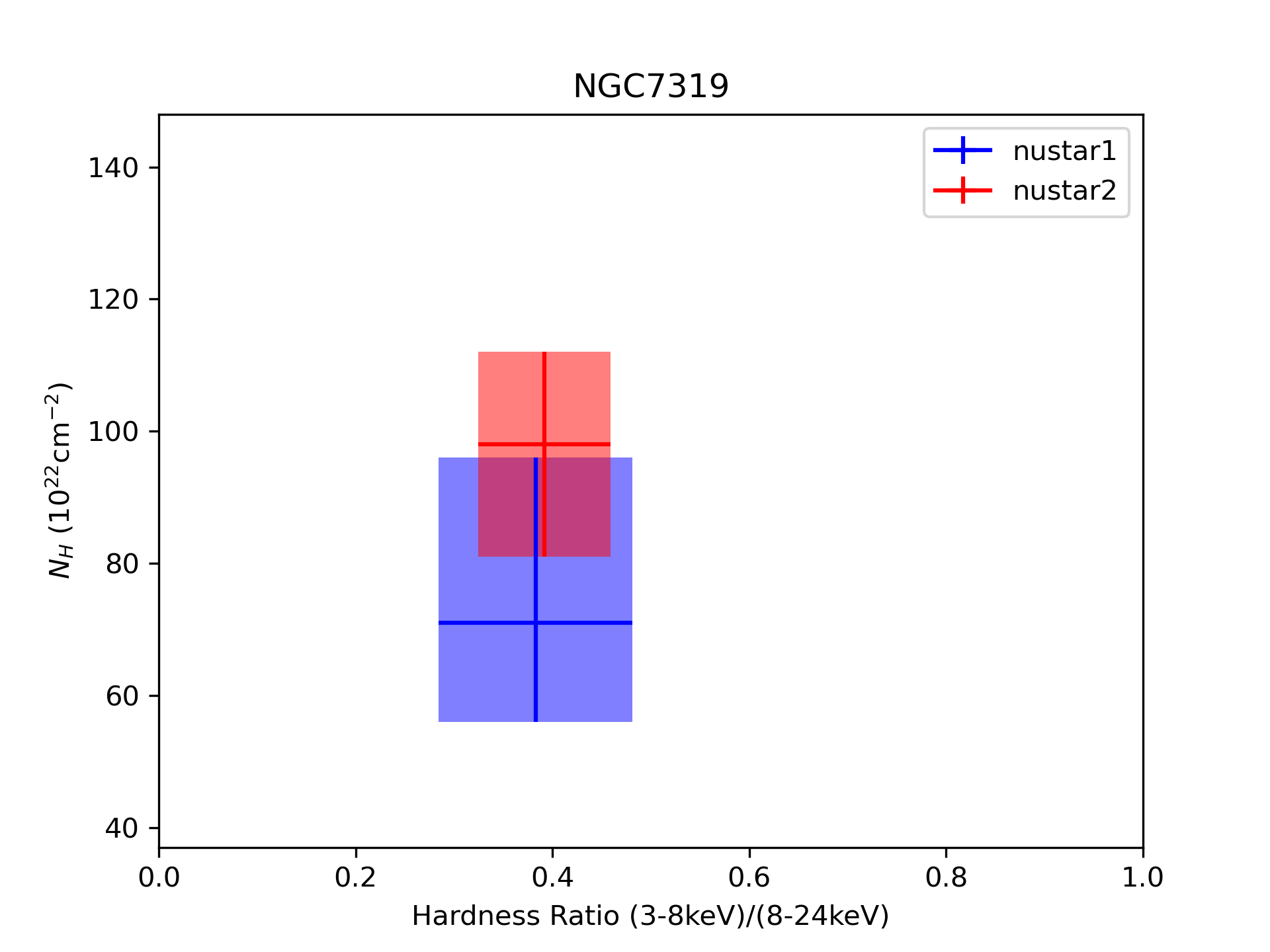}\hfill
    \includegraphics[width=0.33\textwidth]{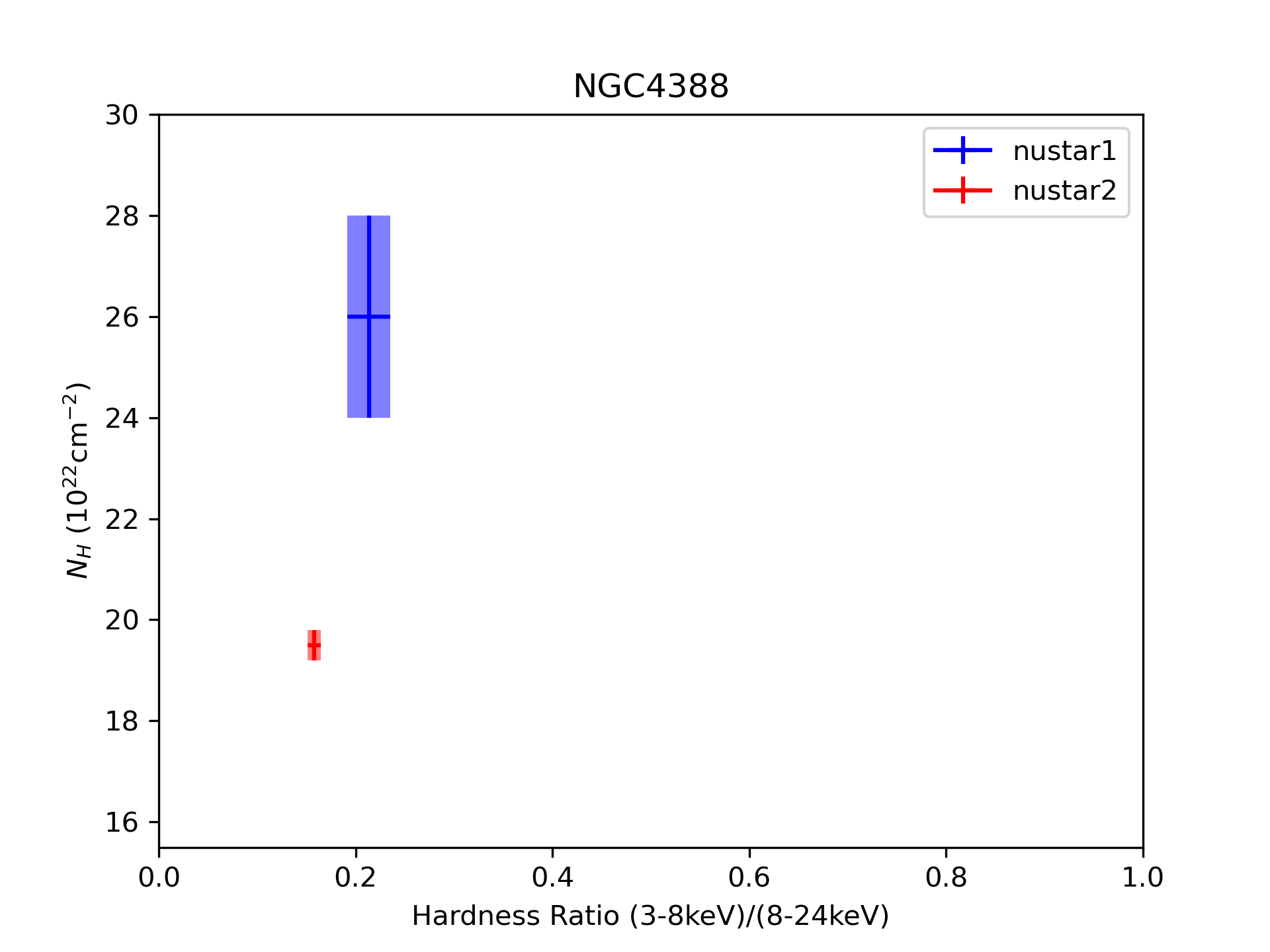}\hfill
    \caption{Direct comparison of the best-fit $N_H$ values from \texttt{UXCLUMPY} to the \nustar\ hardness ratio. As can be seen, the hardness ratio is able to predict $N_H$ variability in NGC 4388 and non-variability in 3C 105 and NGC 7319.}
    \label{fig:nustar_results}
\end{figure*}

The results for the three \nustar\ sources are shown in Figure \ref{fig:nustar_results}. These plots show the best-fit $N_H$ value obtained from the \texttt{UXCLUMPY} model against the single hardness ratio defined in Section \ref{subsection:nustar} for \nustar\ data. As can be seen from the figures, Criterion\,1 correctly predicts the variability in NGC 4388 and also the non-variability in 3C 105 and NGC 7319. 

The predictions are also correct for all three sources when Criterion\,2 is used. For 3C 105, the value for the $\textit{HR}$ fit is $\chi_{\textit{HR}}^2=1.56$ and the the values for the $N_H$ fits are $\chi_{N_H}^2=0.51,0.43,1.25$ for \texttt{UXCLUMPY}, \texttt{MYTorus}, and \texttt{borus02} respectively. Similarly, for NGC 7319, $\chi_{\textit{HR}}^2=0.02$ and $\chi_{N_H}^2=0.92,0.89,1.15$. For the variable source NGC 4388, $\chi_{\textit{HR}}^2=52.4$ and $\chi_{N_H}^2=120,119,94$.

Although the sample size is very small, the \nustar\ hardness ratio seems to be better at predicting $N_H$ variability. It would not be surprising if this is the case, considering the energy bands we are able to use with \nustar\ might be better aligned to detect changes in line-of-sight absorption for $z\sim0$ AGN with moderate obscuration. This could be due to the fact that increasing $N_H$ only significantly affects the 3-8\,keV band which leads to a predictable increase in $\textit{HR}_{nu}$. On the other hand, for $\textit{HR}_1$ and $\textit{HR}_2$, an increase in $N_H$ affects both energy bands differently depending on the amount of absorption and reflection, leading to a less predictable change in the hardness ratios. Of course, we cannot make any definitive statements with only three observations. A larger sample of \nustar\ observations is needed to confirm this.

\section{Summary and Conclusion} \label{sec:Conclusion}

In this work, we introduced a method to predict variability in line-of-sight $N_H$ for an AGN, without having to perform difficult and time-consuming spectral modelling. This would allow the user to quickly sift through many X-ray observations to flag the sources that are most likely to experience $N_H$ variability. These flagged sources can then be studied further by performing a full spectral fitting to obtain accurate $N_H$ values. 

To do this, we used variability in hardness ratio as a proxy for variability in $N_H$. Two different hardness ratios were defined to account for a possible degeneracy in highly obscured scenarios. Two different critera were used to determine whether observations are `variable.' Criterion\,1 considers two observations variable if the 90\,\% confidence intervals are inconsistent with each other. Criterion\,2 considers the $\chi^2$ fit assuming there is no variability.

We tested our prediction method on a sample of 12 sources with $N_H$ values determined through careful spectral modeling, and provided different interpretations of the results. We conclude that our method can be a useful tool for selecting samples of likely $N_H$ variable AGN. Criterion\,1 seems to be a good overall predictor while Criterion\,2 is not as good overall (at a 90\,\% confidence level), but the sensitivity can easily be adjusted to suit the requirements of a particular project, resulting in a very flexible tool. We reiterate, this method is not to be used as a substitute for measuring the $N_H$ via spectral fitting. Rather, it is only an indicator of variability between two observations. 

In a future paper, we will apply this method to a larger sample of sources with unknown $N_H$ values. We will flag the sources with variable $HR$ and study those with careful spectral fitting.

\section{Acknowledgments}

N.T.A., M.A., R.S., A.P. and I.C. acknowledge funding from NASA under contracts 80NSSC19K0531, 80NSSC20K0045 and, 80NSSC20K834. S.M. acknowledges funding from the INAF ``Progetti di Ricerca di Rilevante Interesse Nazionale'' (PRIN), Bando 2019 (project: ``Piercing through the clouds: a multiwavelength study of obscured accretion in nearby supermassive black holes''). The scientific results reported in this article are based on observations made by the X-ray observatories \cha, \nustar, and \xmm, and has made use of the NASA/IPAC Extragalactic Database (NED), which is operated by the Jet Propulsion Laboratory, California Institute of Technology under contract with NASA. We acknowledge the use of the software package HEASoft.

\bibliography{references}{}
\bibliographystyle{aa}

\appendix

\section{Sample Results}\label{sec:sample_results}

\begin{figure*}[h!]
    \centering
    \includegraphics[width=0.50\textwidth]{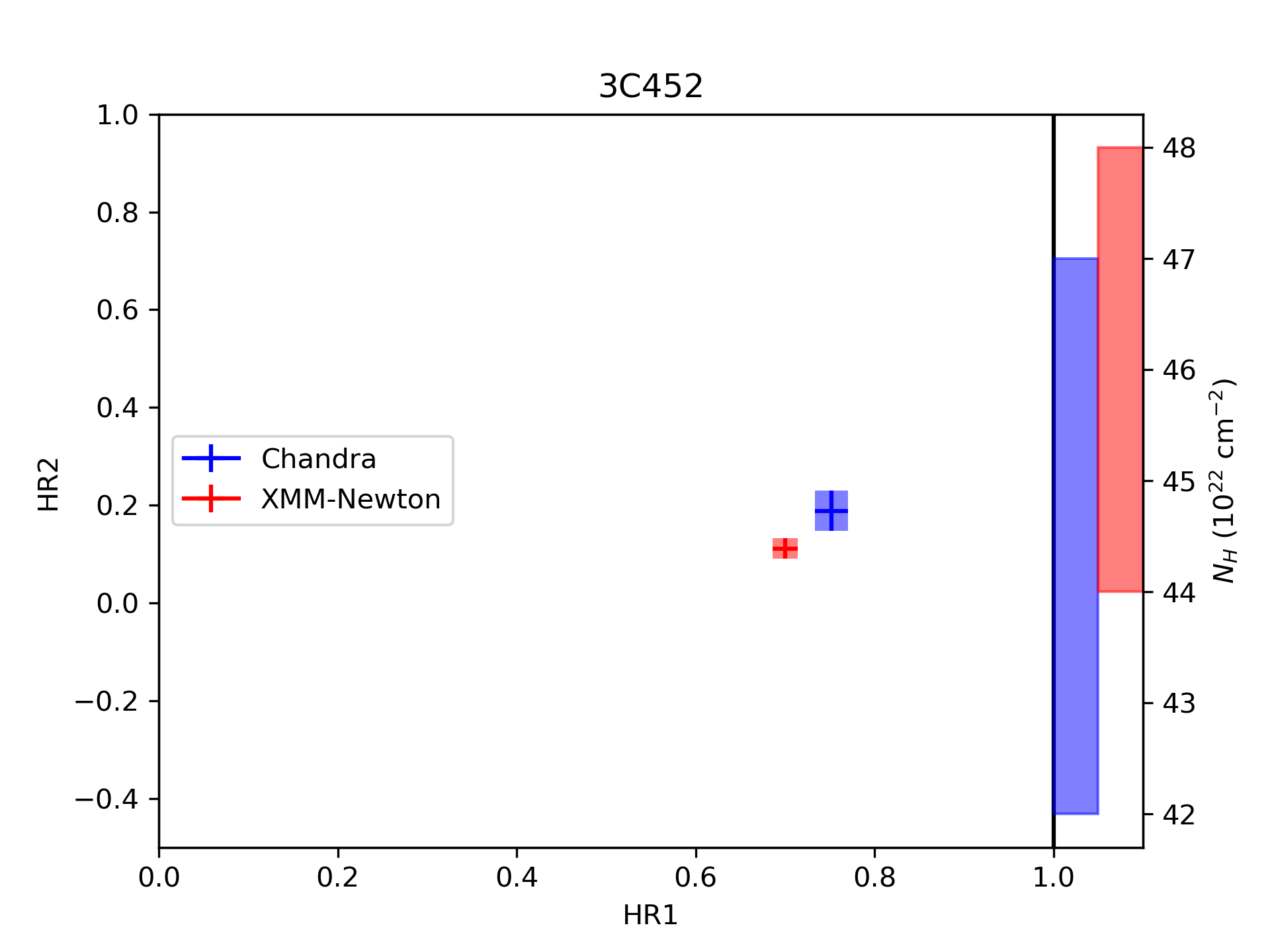}\hfill
    \includegraphics[width=0.50\textwidth]{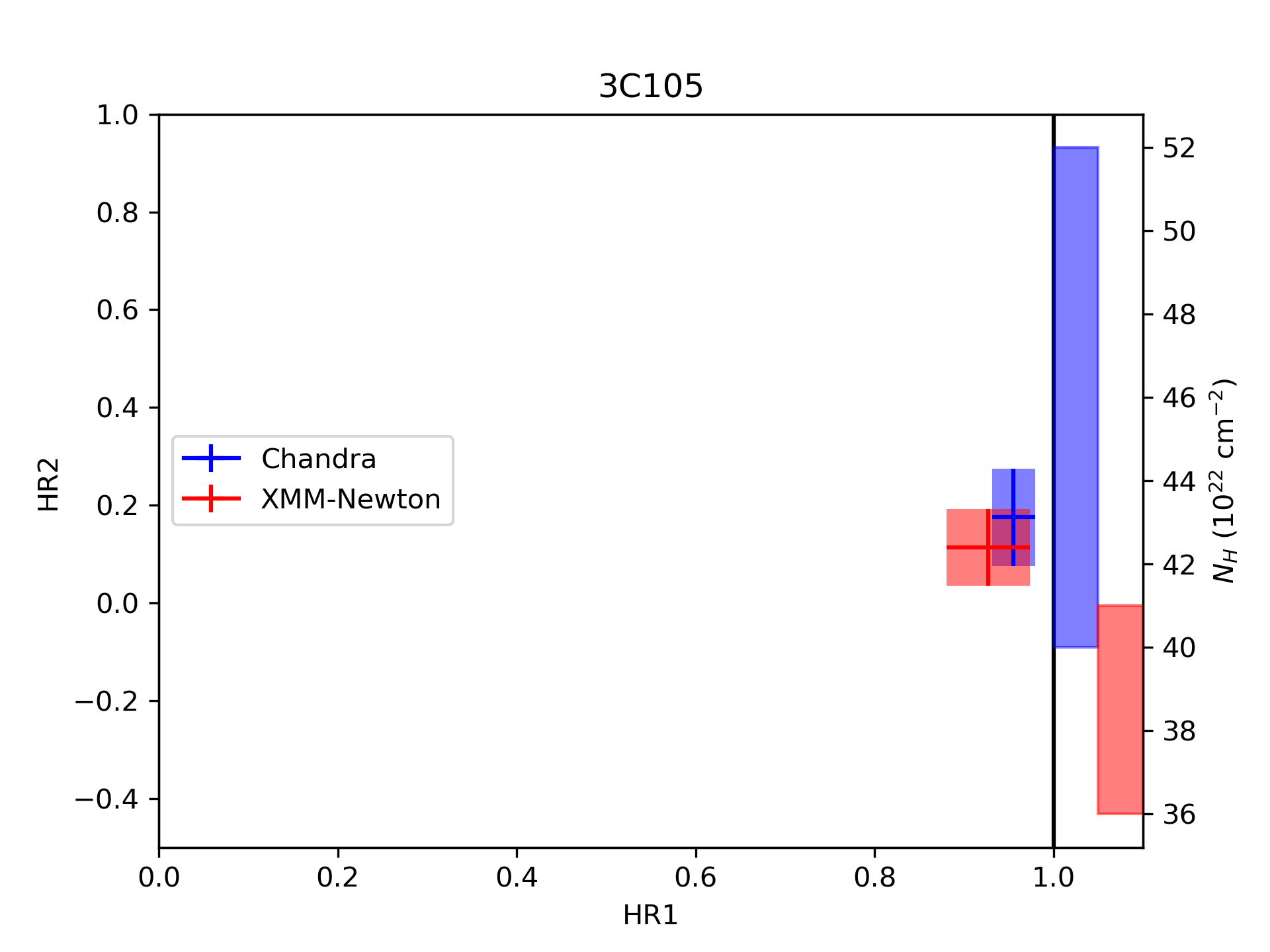}\hfill
    \caption{Results for 3C 452 and 3C 105. The boxes represent the 90\% confidence interval for $\textit{HR}_1$ and $\textit{HR}_2$. The bar on the right shows the 90\% confidence interval for the modeled $N_{H,los}$ with \texttt{UXCLUMPY}, which we take to be the ``true" column density.}
    \label{fig:A1}
\end{figure*}

\begin{figure*}[h!]
    \centering
    \includegraphics[width=0.50\textwidth]{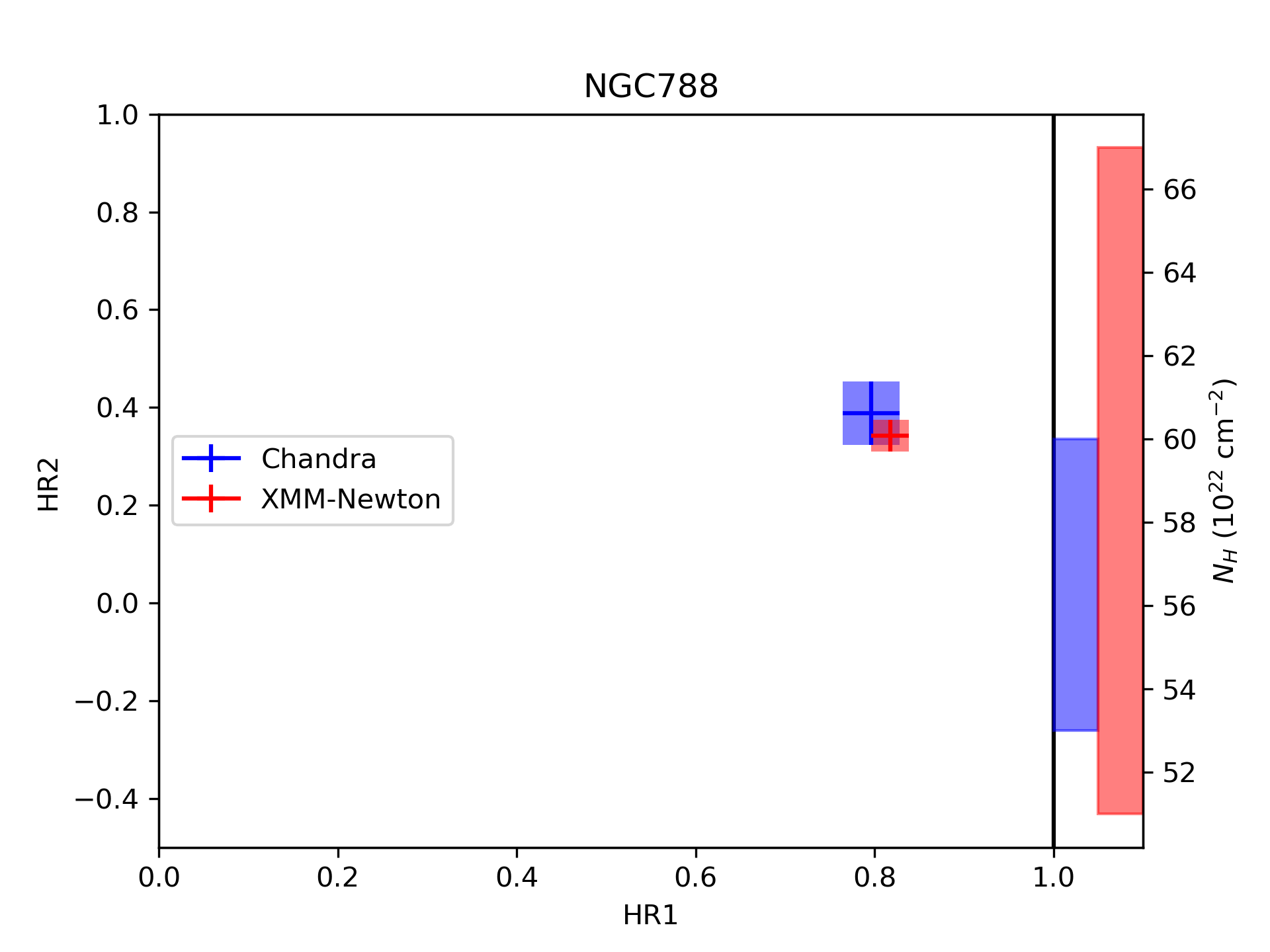}\hfill
    \includegraphics[width=0.50\textwidth]{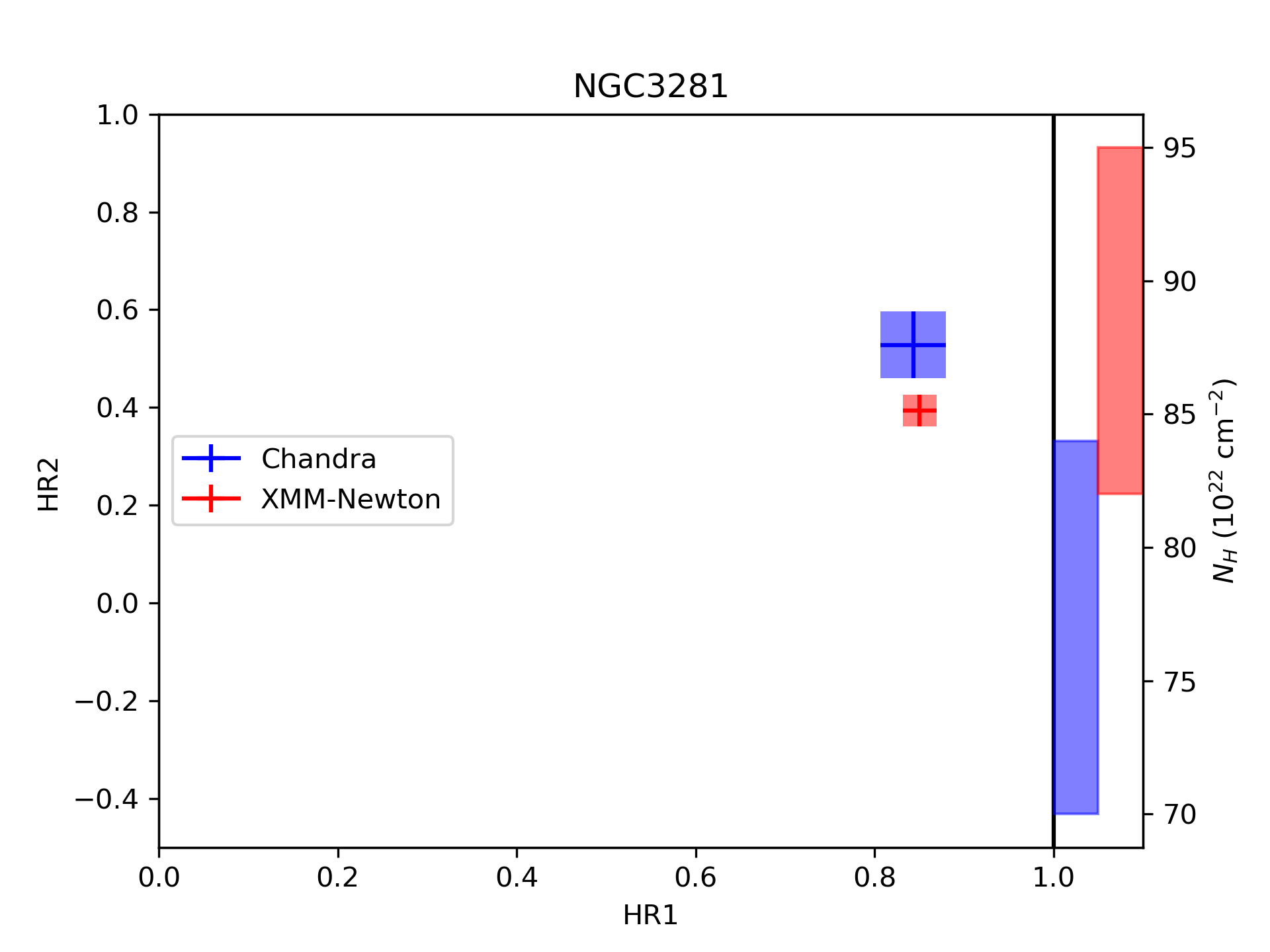}\hfill
    \caption{Results for NGC 788 and NGC 3281. The boxes represent the 90\% confidence interval for $\textit{HR}_1$ and $\textit{HR}_2$. The bar on the right shows the 90\% confidence interval for the modeled $N_{H,los}$ with \texttt{UXCLUMPY}, which we take to be the ``true" column density.}
    \label{fig:A2}
\end{figure*}

\begin{figure*}[h!]
    \centering
    \includegraphics[width=0.50\textwidth]{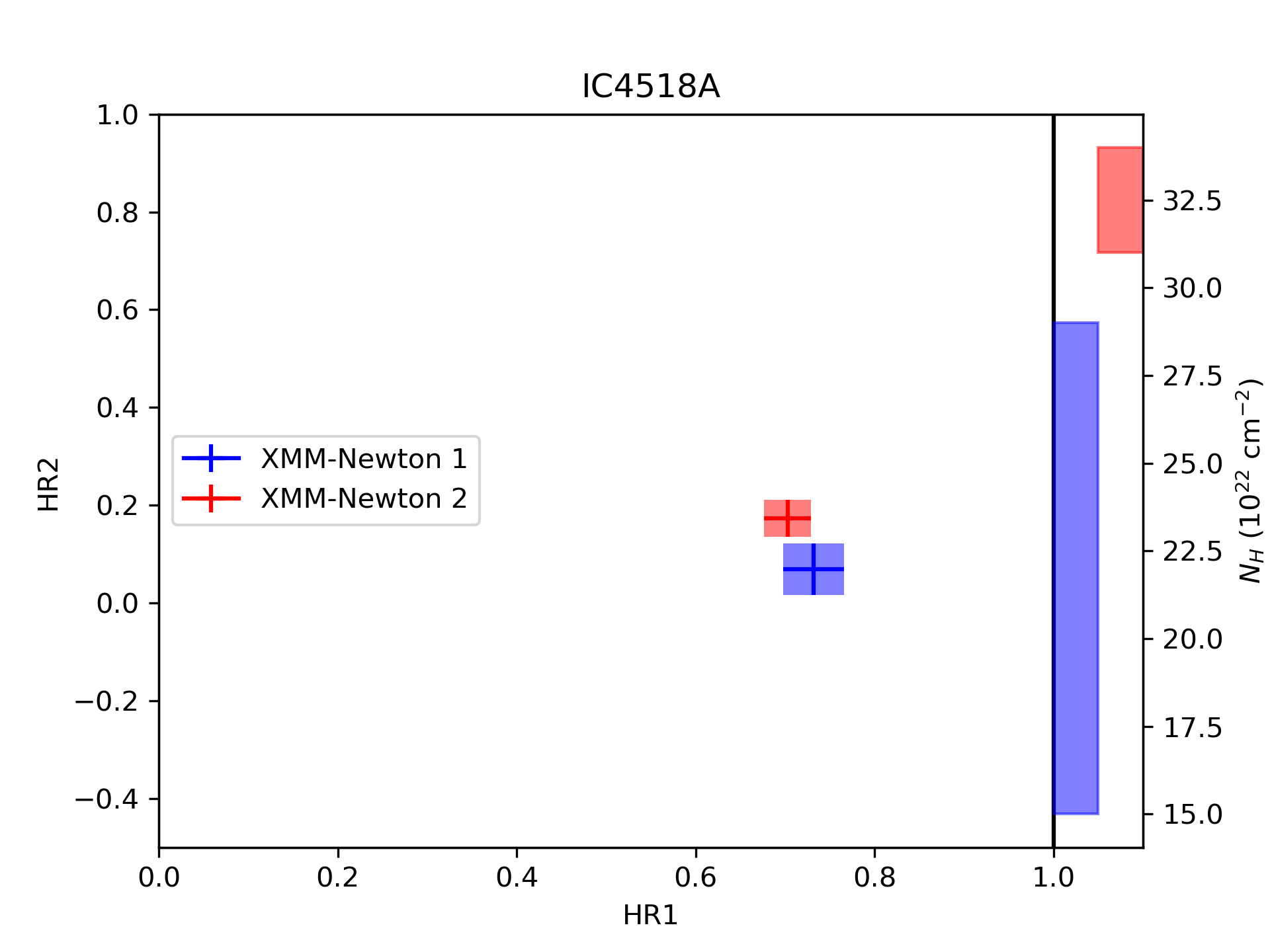}\hfill
    \includegraphics[width=0.50\textwidth]{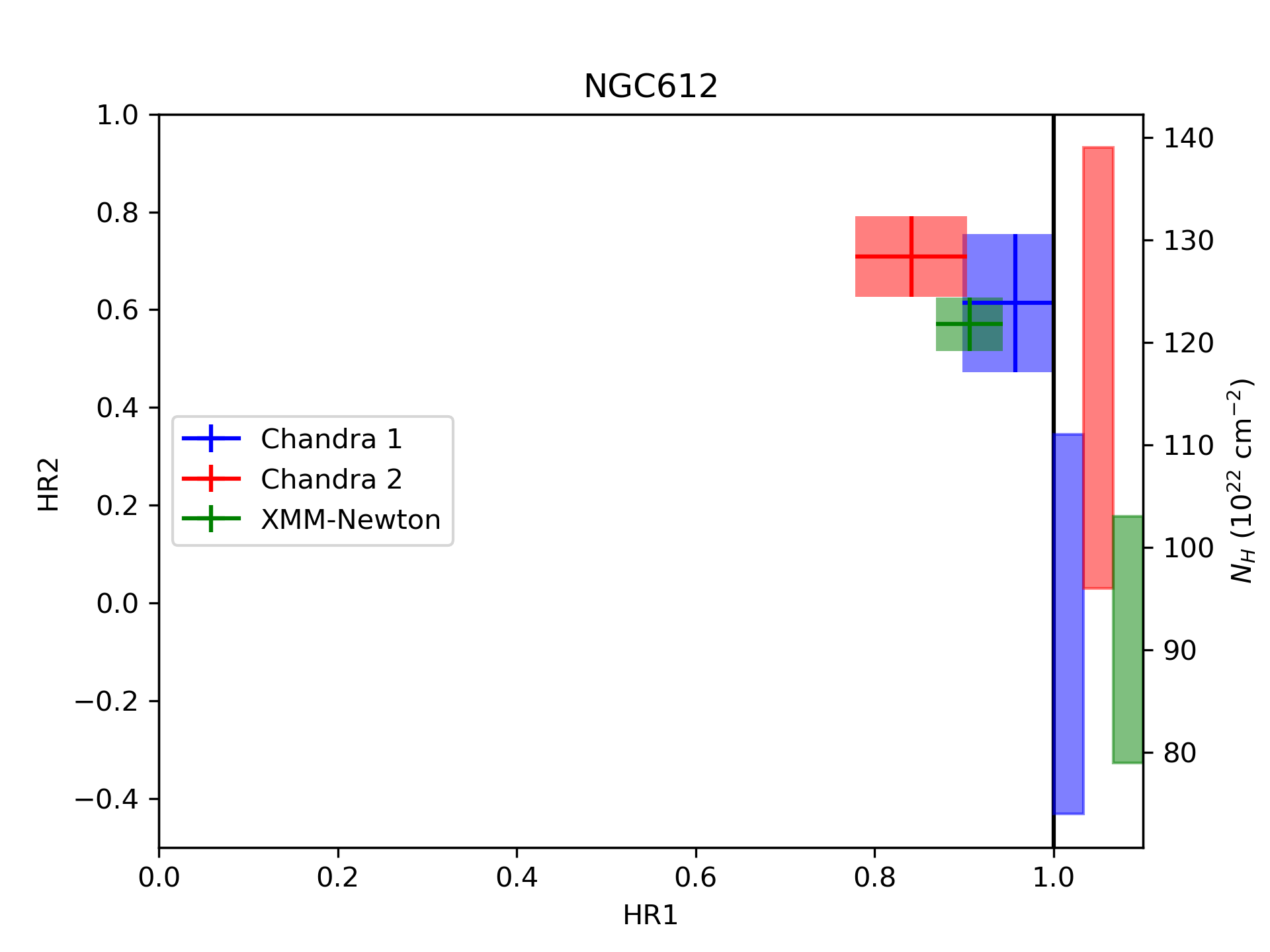}\hfill
    \caption{Results for IC 4518A and NGC 612. The boxes represent the 90\% confidence interval for $\textit{HR}_1$ and $\textit{HR}_2$. The bar on the right shows the 90\% confidence interval for the modeled $N_{H,los}$ with \texttt{UXCLUMPY}, which we take to be the ``true" column density.}
    \label{fig:A3}
\end{figure*}

\begin{figure*}[h!]
    \centering
    \includegraphics[width=0.50\textwidth]{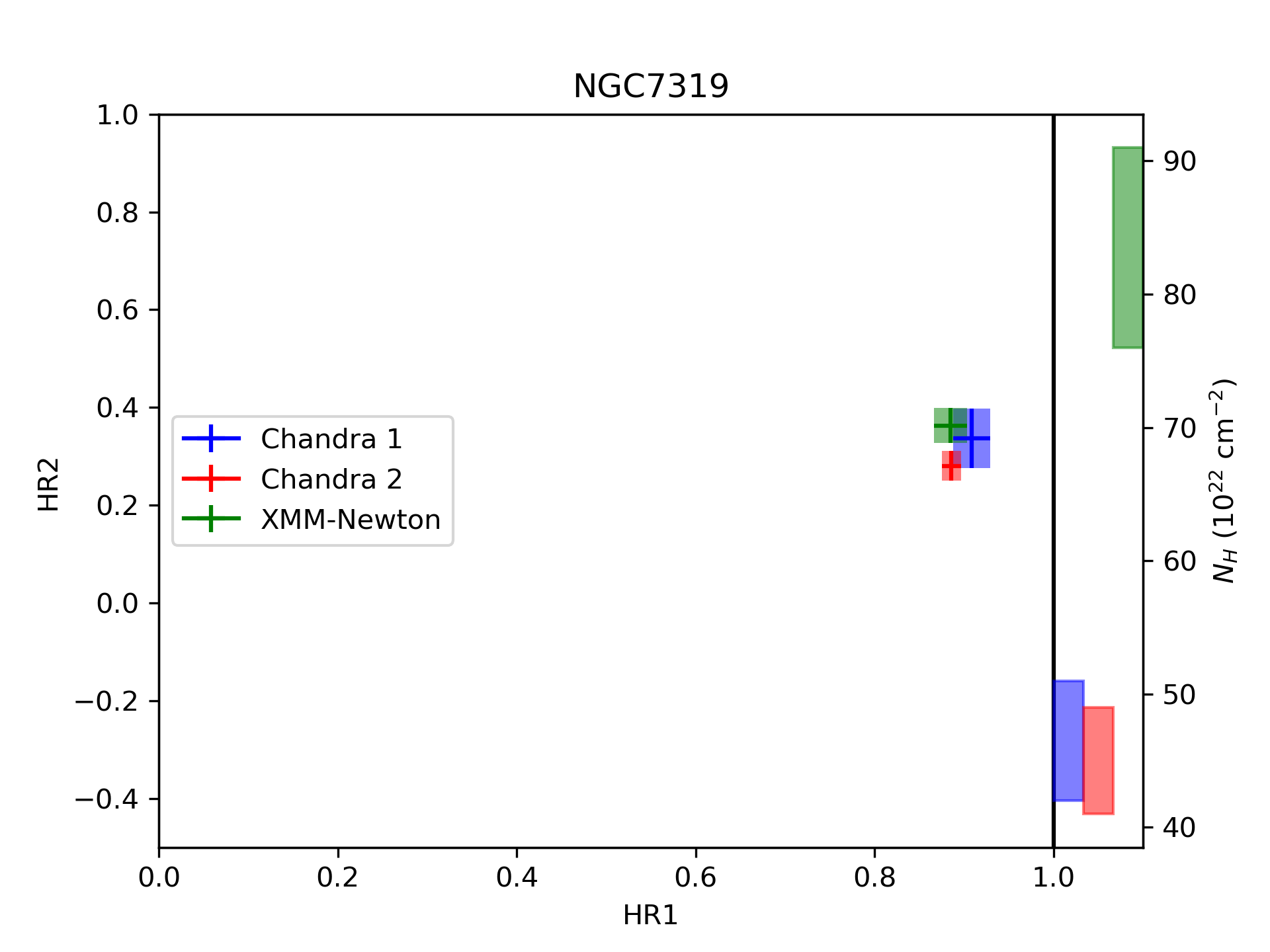}\hfill
    \includegraphics[width=0.50\textwidth]{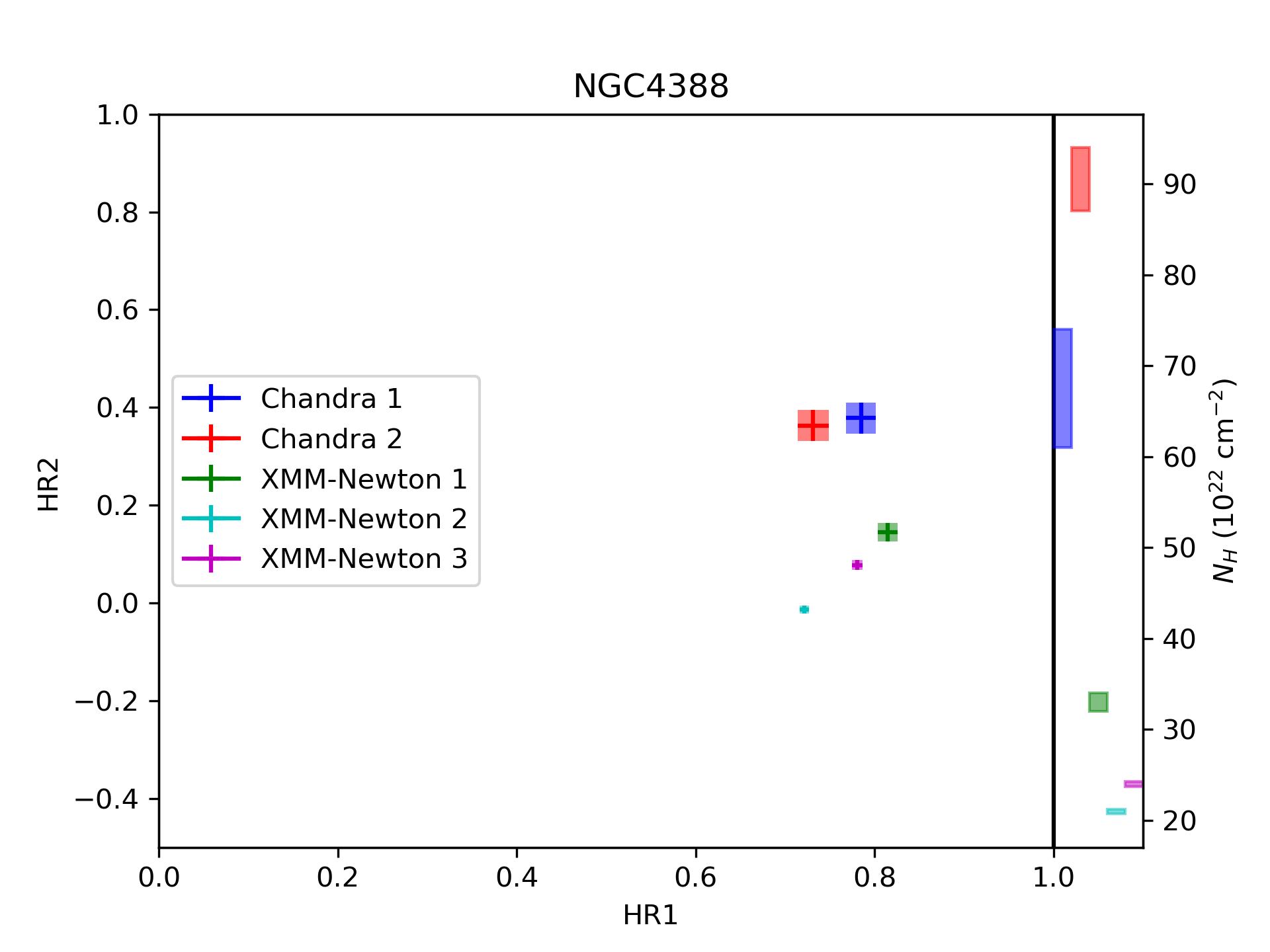}\hfill
    \caption{Results for NGC 7319 and NGC 4388. The boxes represent the 90\% confidence interval for $\textit{HR}_1$ and $\textit{HR}_2$. The bar on the right shows the 90\% confidence interval for the modeled $N_{H,los}$ with \texttt{UXCLUMPY}, which we take to be the ``true" column density.}
    \label{fig:A4}
\end{figure*}

\begin{figure*}[h!]
    \centering
    \includegraphics[width=0.50\textwidth]{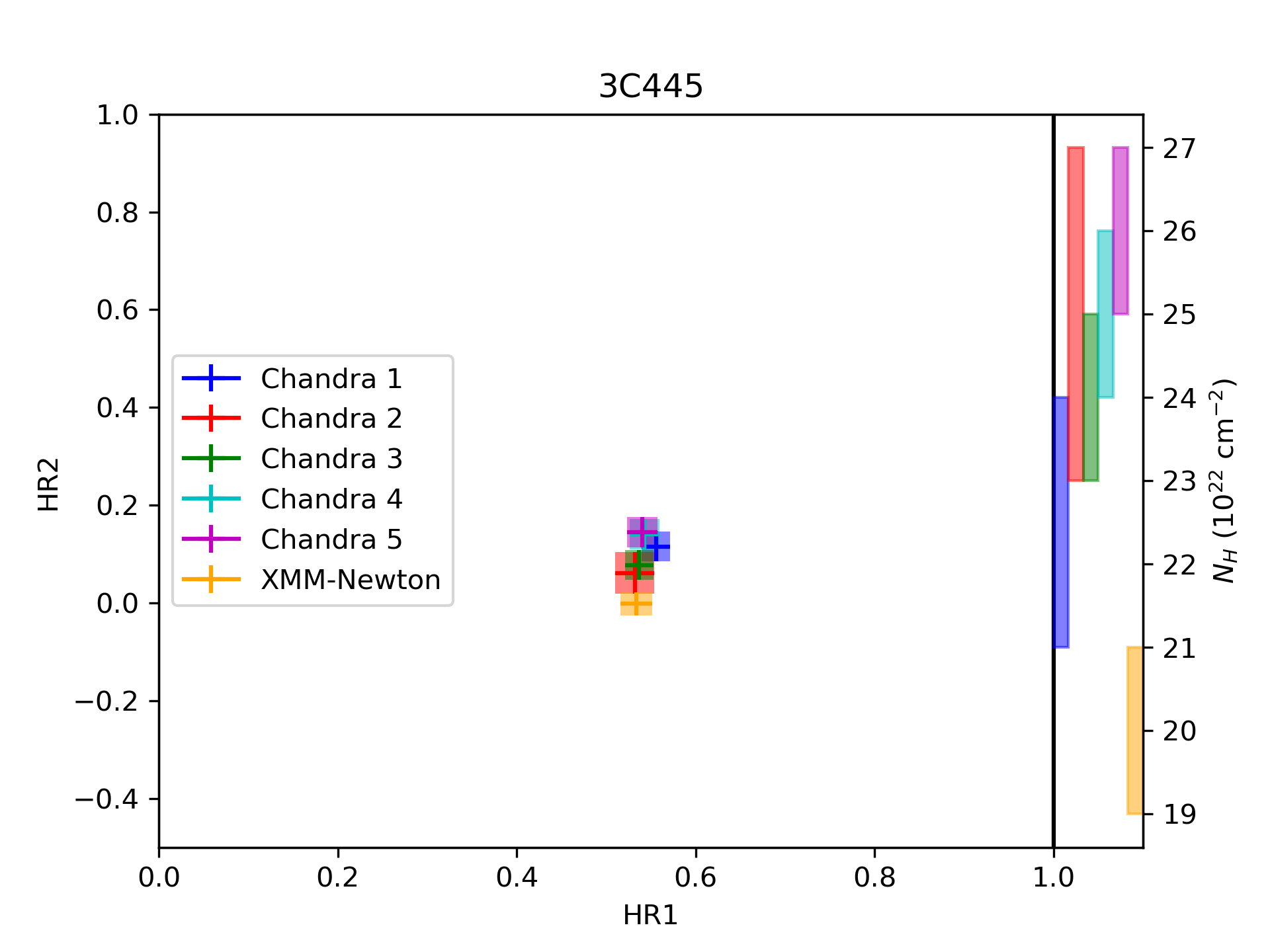}\hfill
    \includegraphics[width=0.50\textwidth]{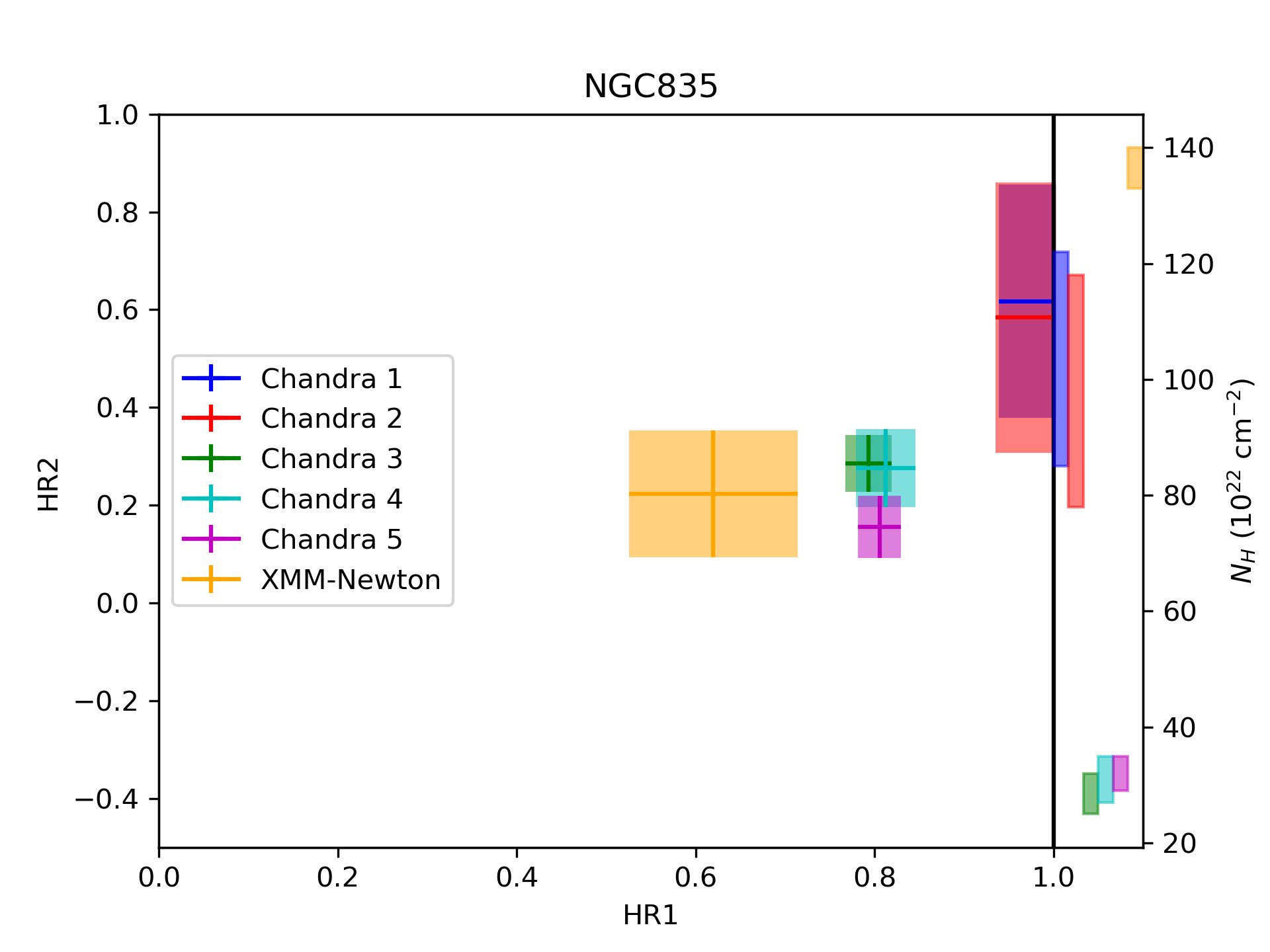}\hfill
    \caption{Results for 3C 445 and NGC 835. The boxes represent the 90\% confidence interval for $\textit{HR}_1$ and $\textit{HR}_2$. The bar on the right shows the 90\% confidence interval for the modeled $N_{H,los}$ with \texttt{UXCLUMPY}, which we take to be the ``true" column density.}
    \label{fig:A5}
\end{figure*}

\begin{figure*}[h!]
    \centering
    \includegraphics[width=0.50\textwidth]{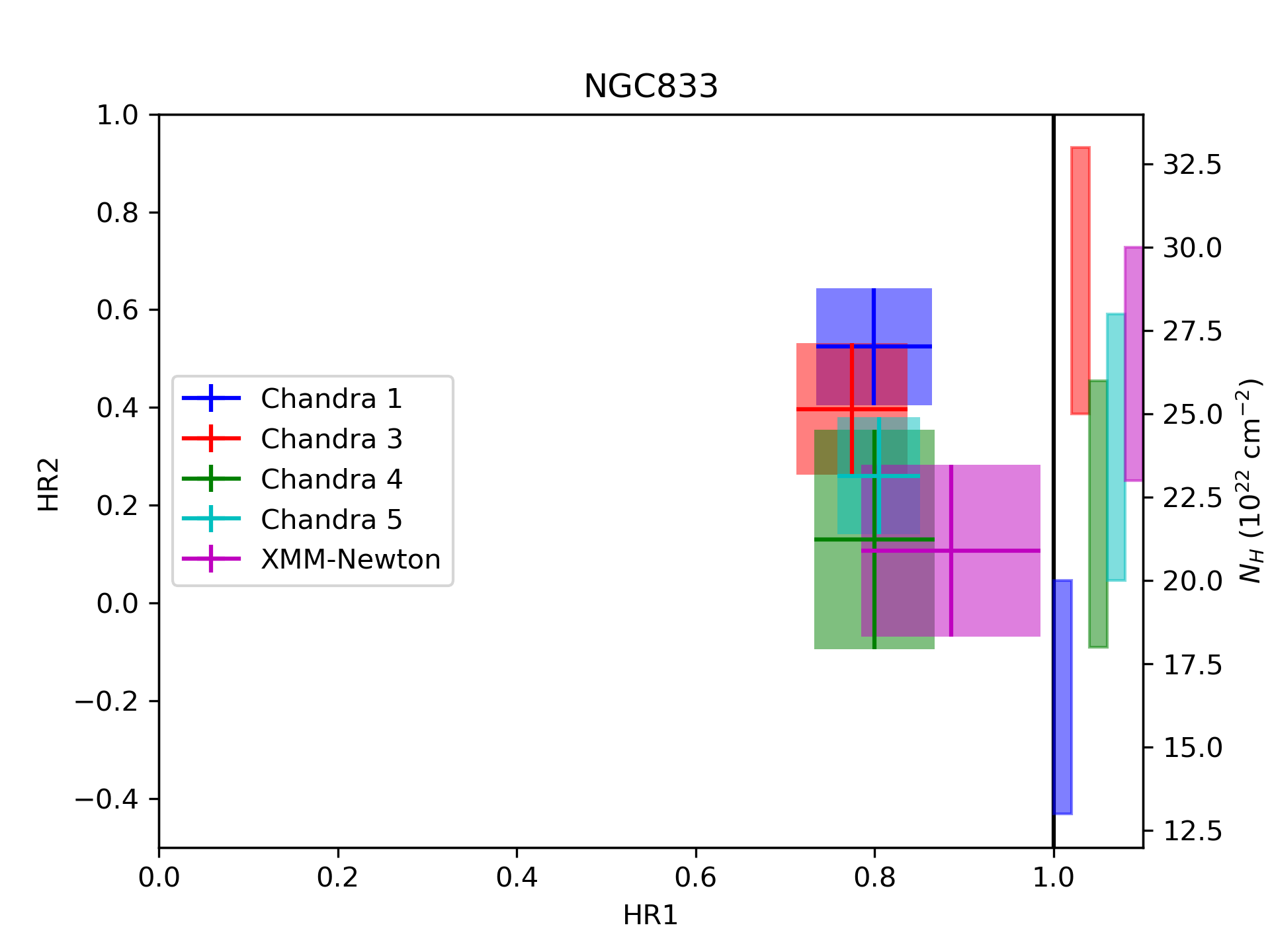}\hfill
    \includegraphics[width=0.50\textwidth]{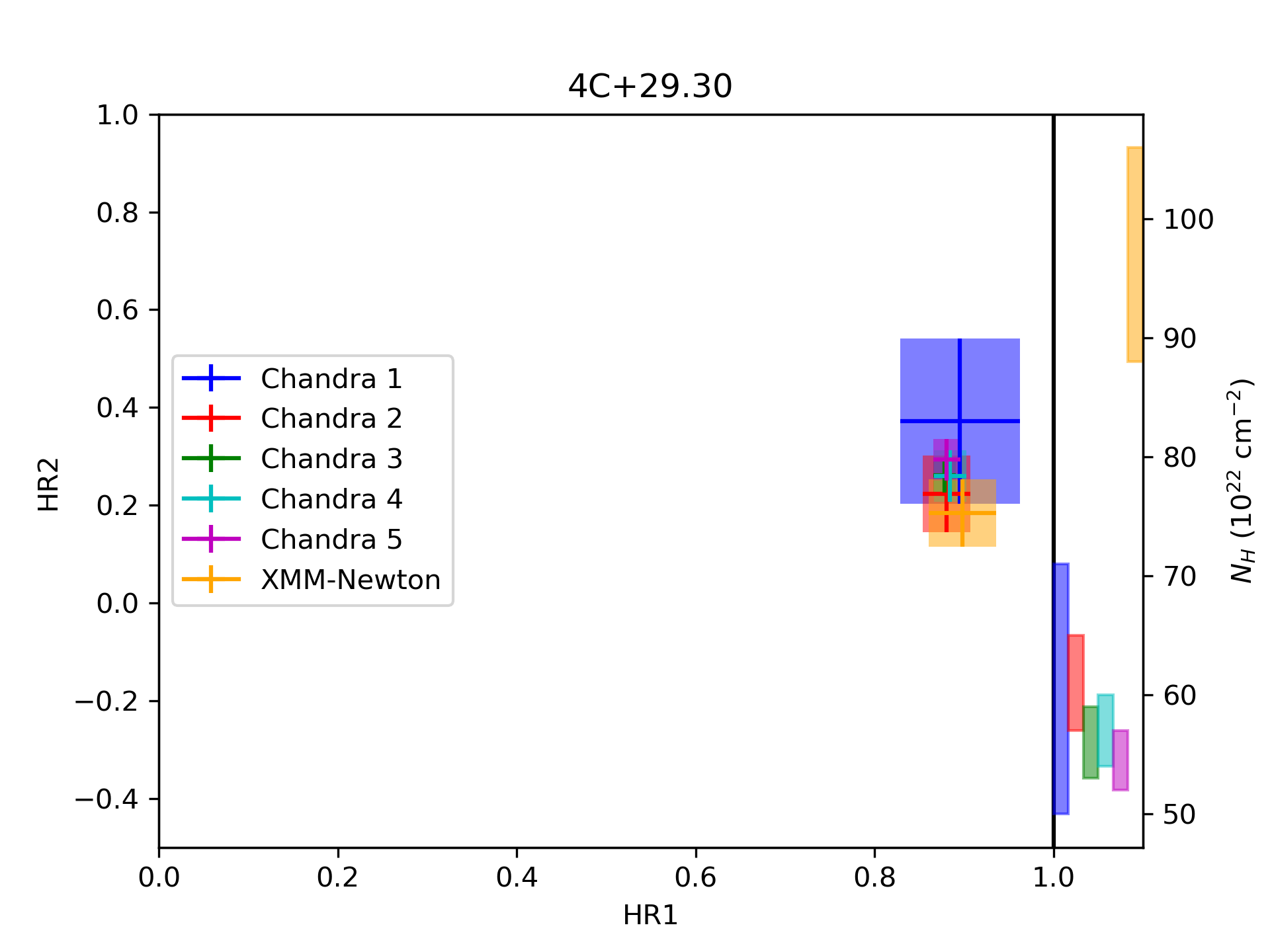}\hfill
    \caption{Results for NGC 833 and 4C+29.30. The boxes represent the 90\% confidence interval for $\textit{HR}_1$ and $\textit{HR}_2$. The bar on the right shows the 90\% confidence interval for the modeled $N_{H,los}$ with \texttt{UXCLUMPY}, which we take to be the ``true" column density.}
    \label{fig:A6}
\end{figure*}

Figures \ref{fig:A1}$-$\ref{fig:A6} show the analysis results for all 12 sources in our sample.

%\begin{figure}
%    \centering
%    \includegraphics[width=0.25\textwidth]{Sources_uxclumpy/3C105.png}\hfill
%    \includegraphics[width=0.25\textwidth]{Sources_uxclumpy/3C445.png}\hfill
%    \includegraphics[width=0.25\textwidth]{Sources_uxclumpy/3C452.png}\hfill
%    \includegraphics[width=0.25\textwidth]{Sources_uxclumpy/4C+29.30.png}\hfill
%    \includegraphics[width=0.25\textwidth]{Sources_uxclumpy/IC4518A.png}\hfill
%    \includegraphics[width=0.25\textwidth]{Sources_uxclumpy/NGC3281.png}\hfill
%    \includegraphics[width=0.25\textwidth]{Sources_uxclumpy/NGC4388.png}\hfill
%    \includegraphics[width=0.25\textwidth]{Sources_uxclumpy/NGC612.png}\hfill
%    \includegraphics[width=0.25\textwidth]{Sources_uxclumpy/NGC7319.png}\hfill
%    \includegraphics[width=0.25\textwidth]{Sources_uxclumpy/NGC788.png}\hfill
%    \includegraphics[width=0.25\textwidth]{Sources_uxclumpy/NGC833.png}\hfill
%    \includegraphics[width=0.25\textwidth]{Sources_uxclumpy/NGC835.png}\hfill
%    \caption{Results all sources in our sample. The errorboxes represent the 90\% confidence interval for $\textit{HR}_1$ and $\textit{HR}_2$. The bar on the right shows the 90\% confidence interval for the modeled $N_{H,los}$ which we take to be the ``true" column density.}
%    \label{fig:all_results}
%\end{figure}

%\newpage

\section{Confusion Matrices}\label{sec:confusion_matrices}

\begin{figure*}[h!]
    \centering
    \includegraphics[width=0.33\textwidth]{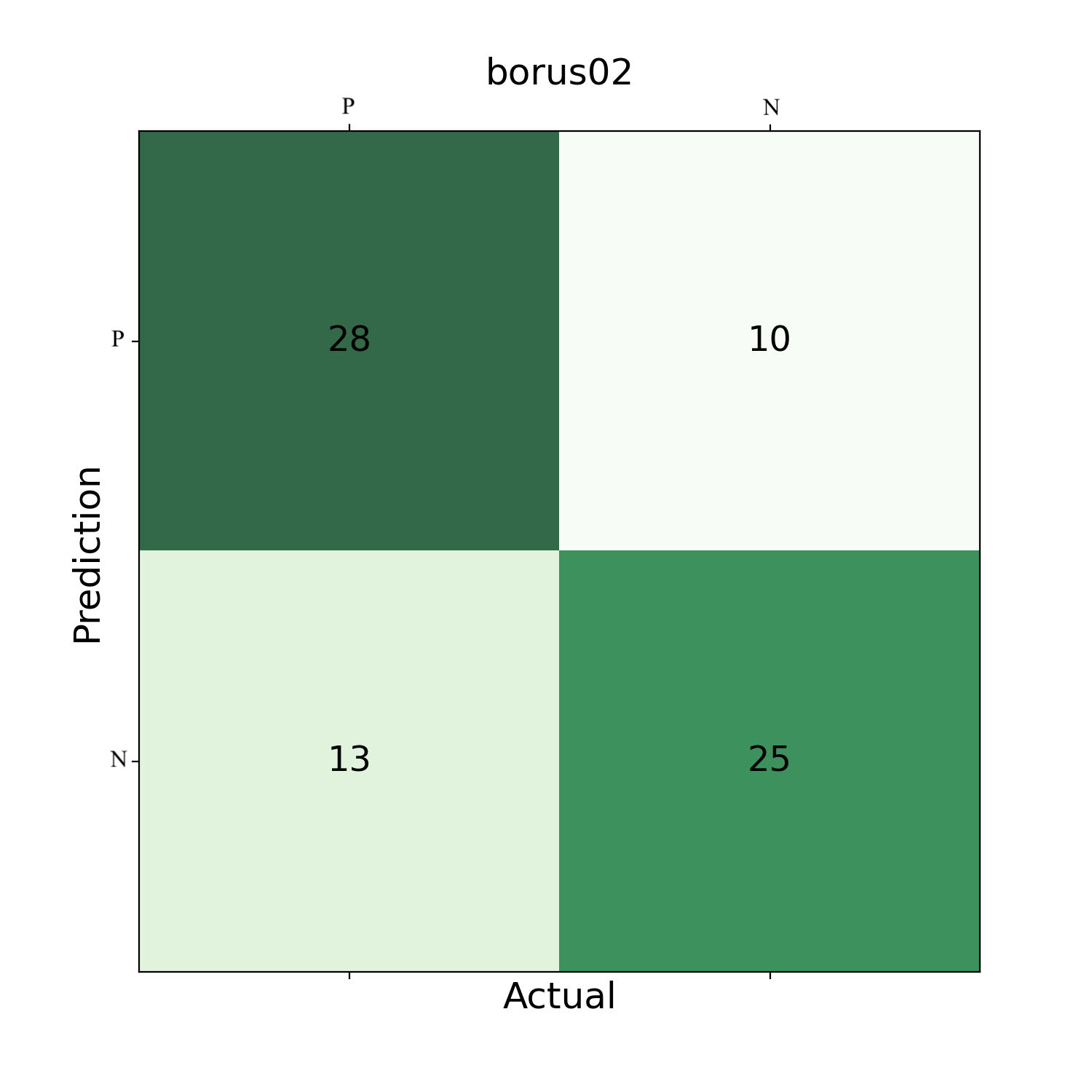}\hfill
    \includegraphics[width=0.33\textwidth]{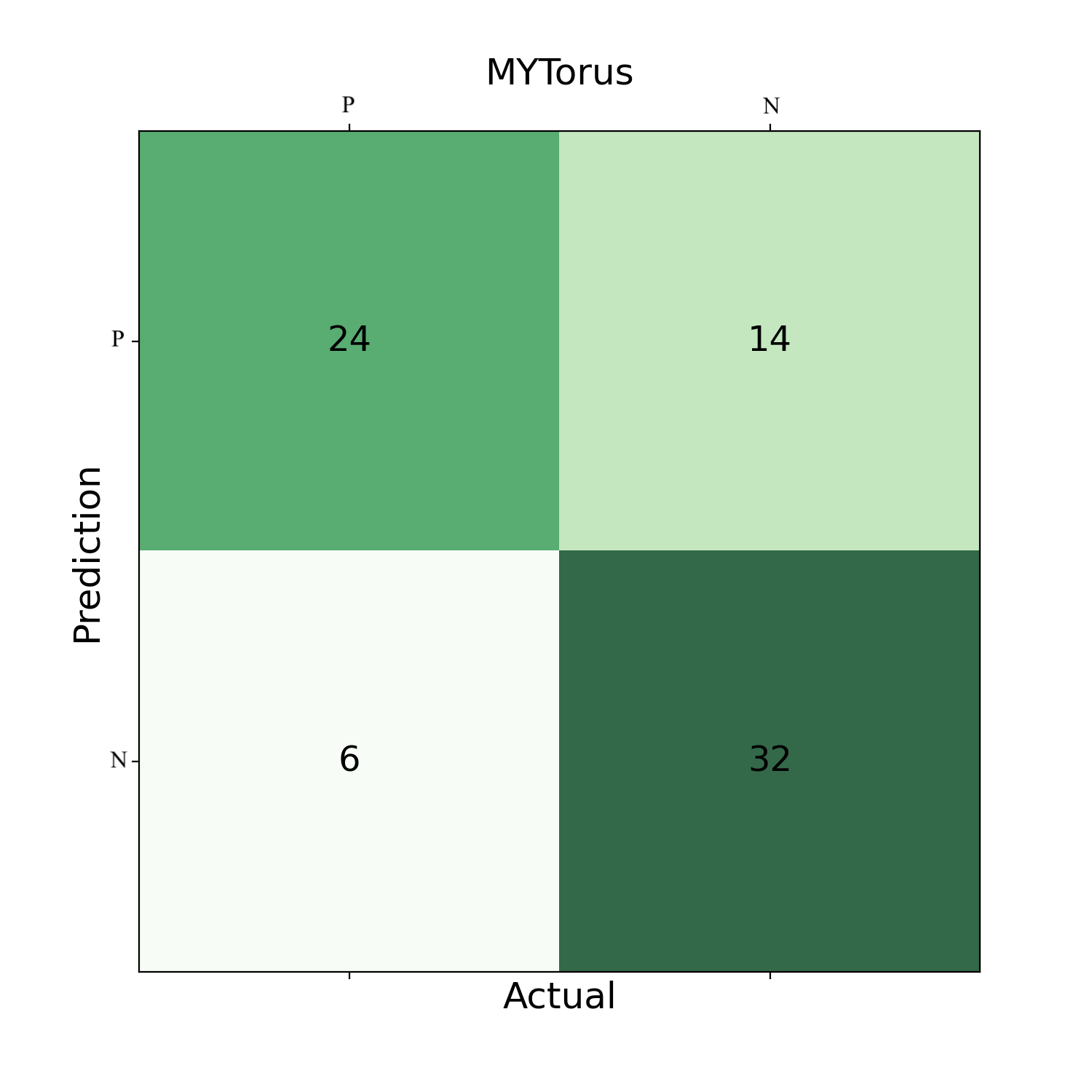}\hfill
    \includegraphics[width=0.33\textwidth]{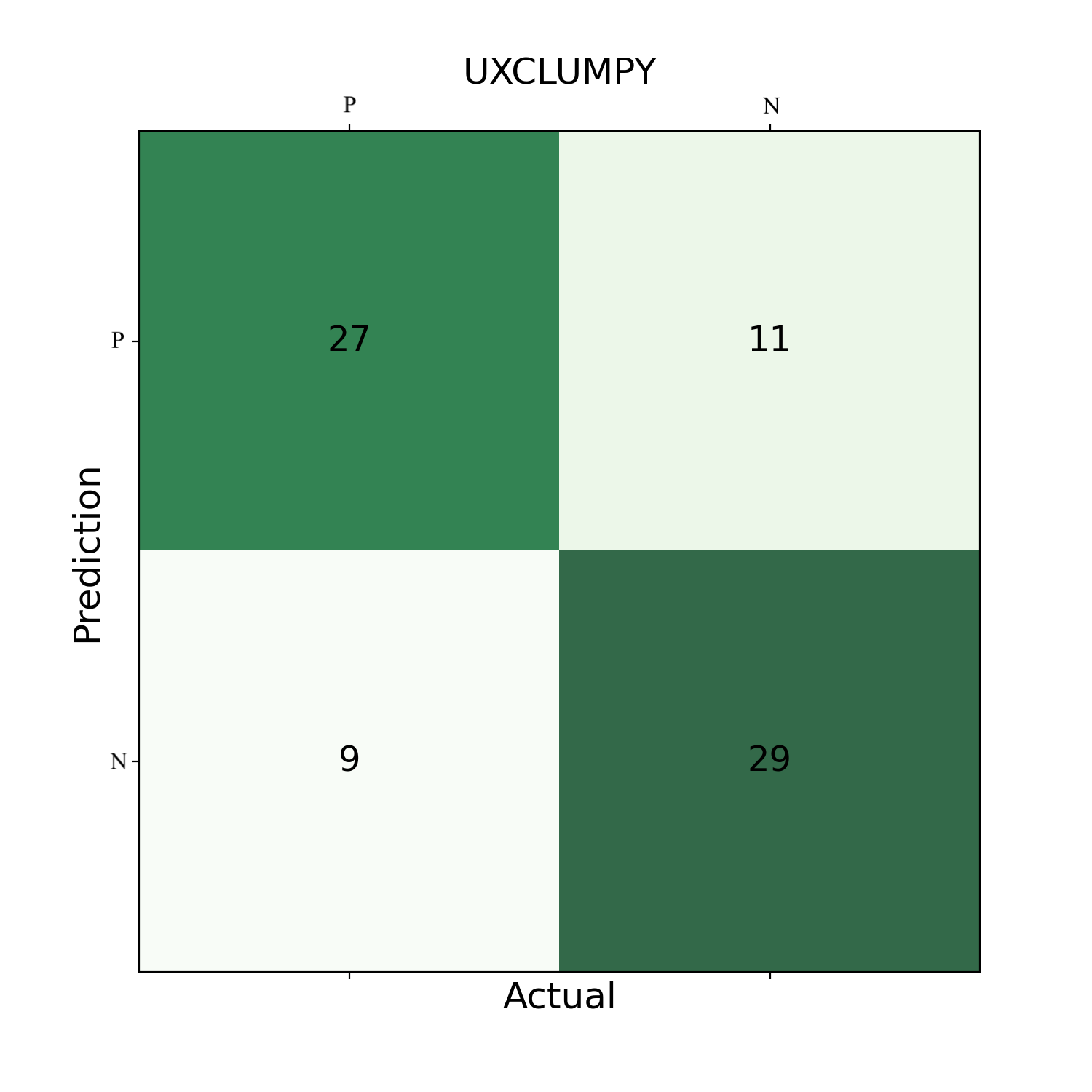}\hfill
    \caption{Confusion matrices for all three models using Criterion\,1 (the overlap method) considering $\textit{HR}_1$\&$\textit{HR}_2$. These show that the method is fairly good at classifying sources as variable or not variable.}
    \label{fig:B1}
\end{figure*}

\begin{figure*}[h!]
    \centering
    \includegraphics[width=0.33\textwidth]{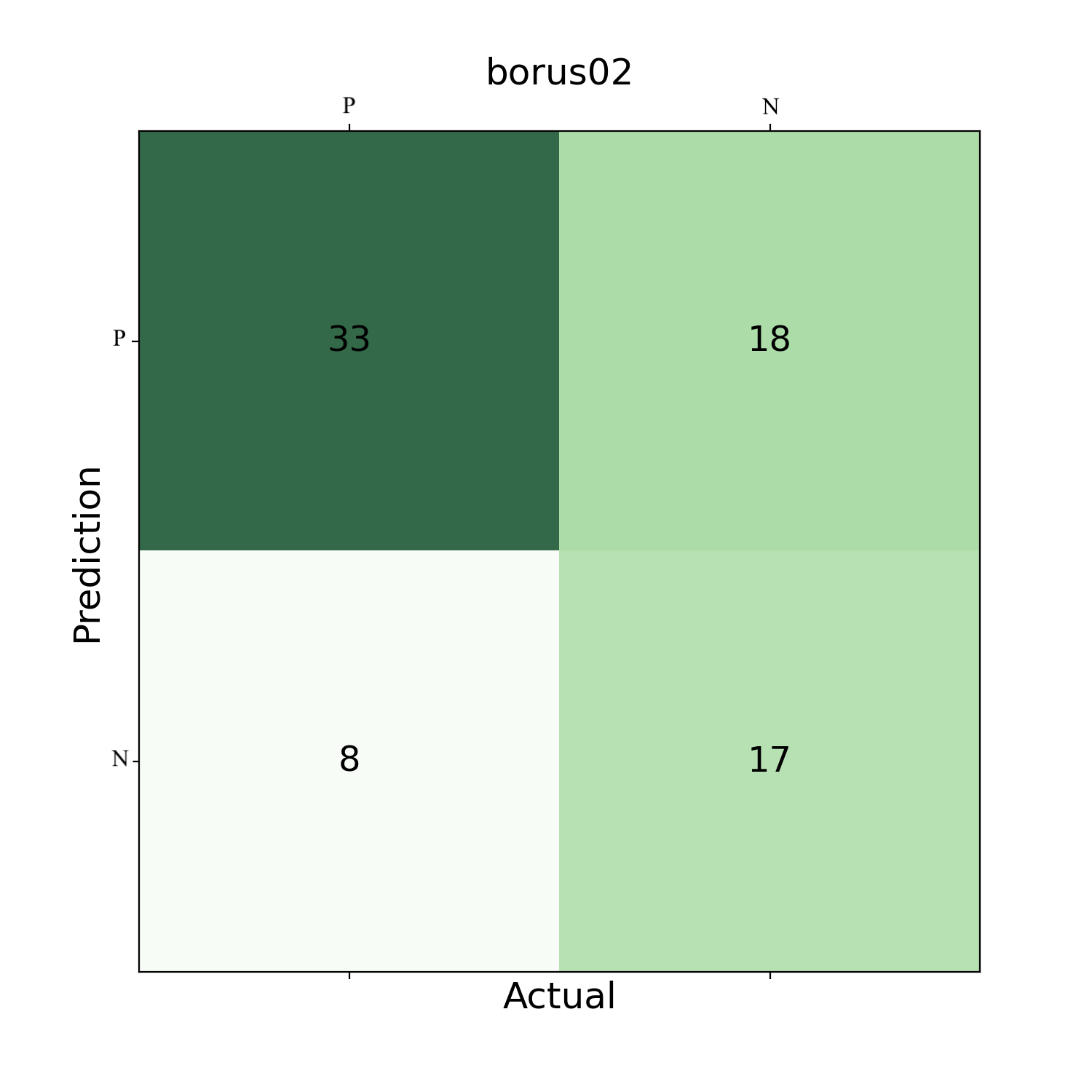}\hfill
    \includegraphics[width=0.33\textwidth]{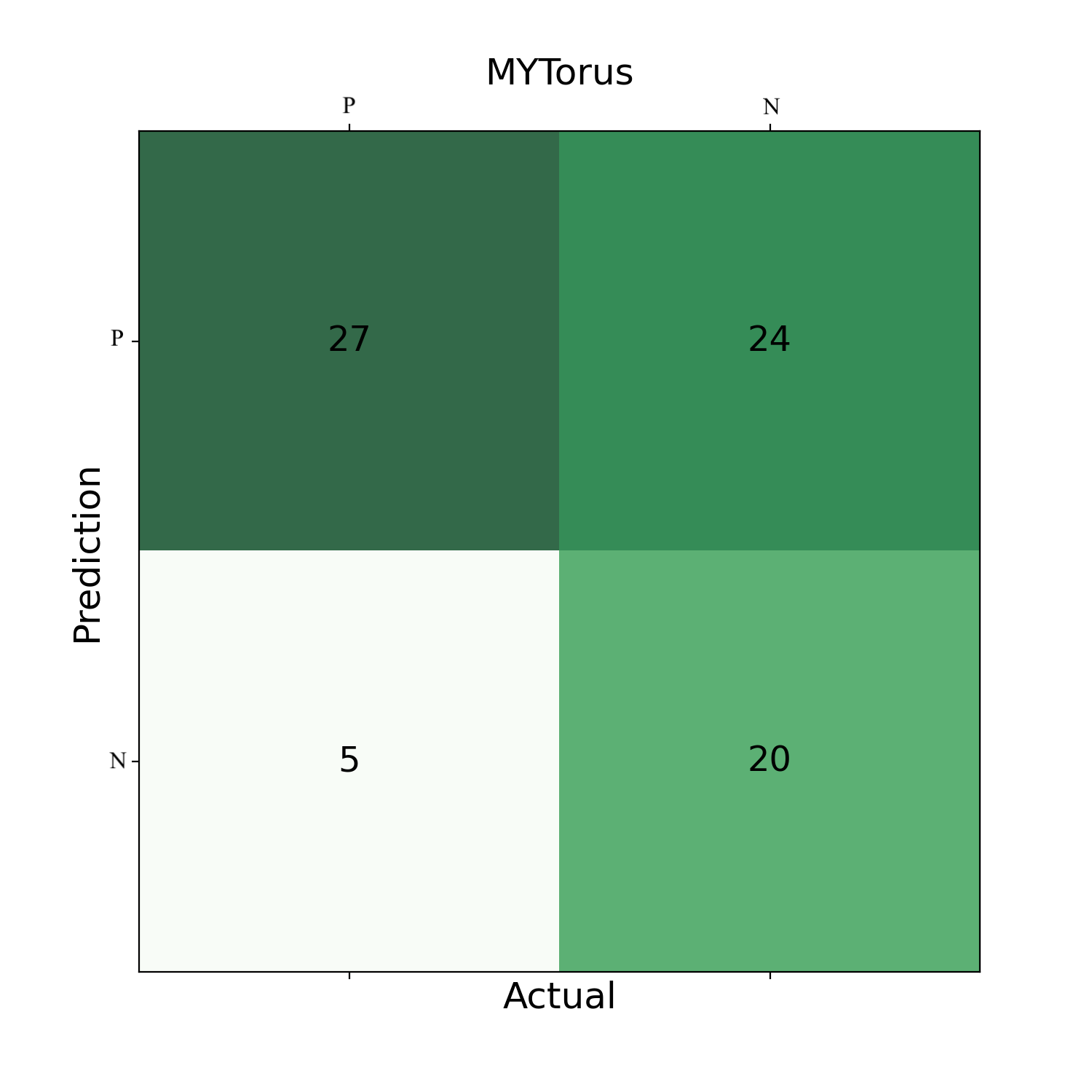}\hfill
    \includegraphics[width=0.33\textwidth]{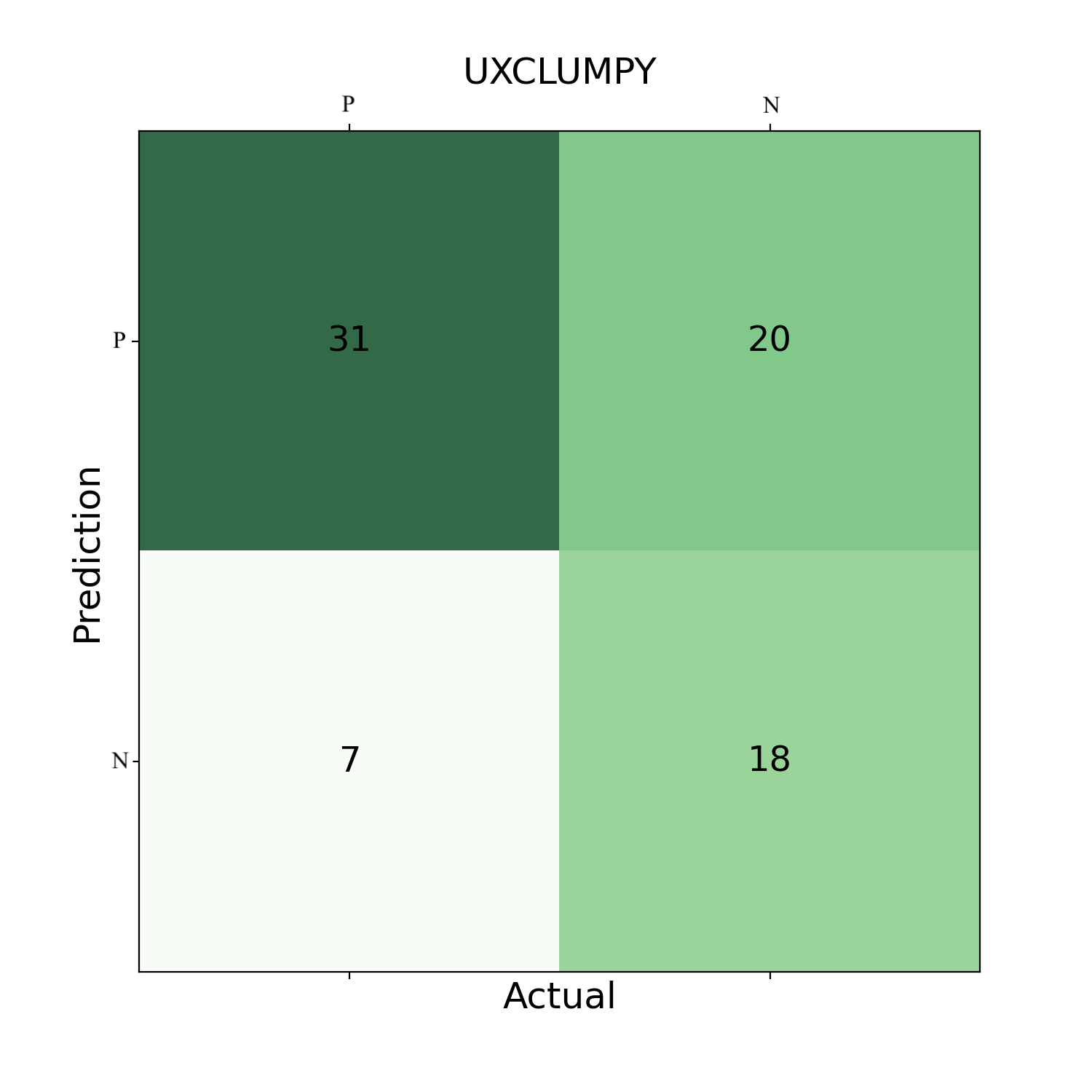}\hfill
    \caption{Confusion matrices for all three models using Criterion\,2 at a 90\,\% confidence level considering $\textit{HR}_1$\&$\textit{HR}_2$. These show that the method is good at avoiding false negatives meaning that most of the variable sources in a sample will be flagged. However, this comes at the expense of flagging as variable more sources that are not variable (top right).}
    \label{fig:B2}
\end{figure*}

\begin{figure*}[h!]
    \centering
    \includegraphics[width=0.33\textwidth]{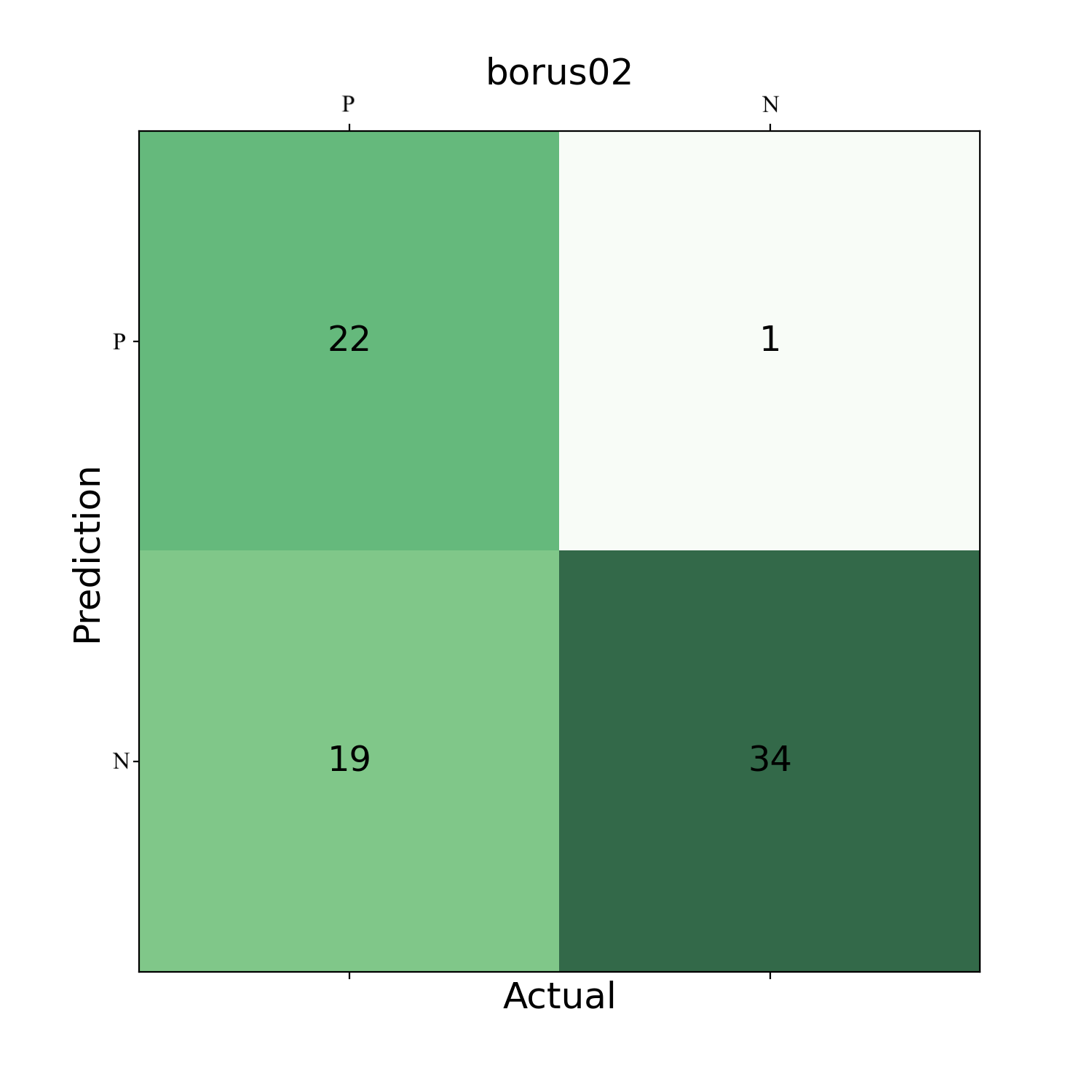}\hfill
    \includegraphics[width=0.33\textwidth]{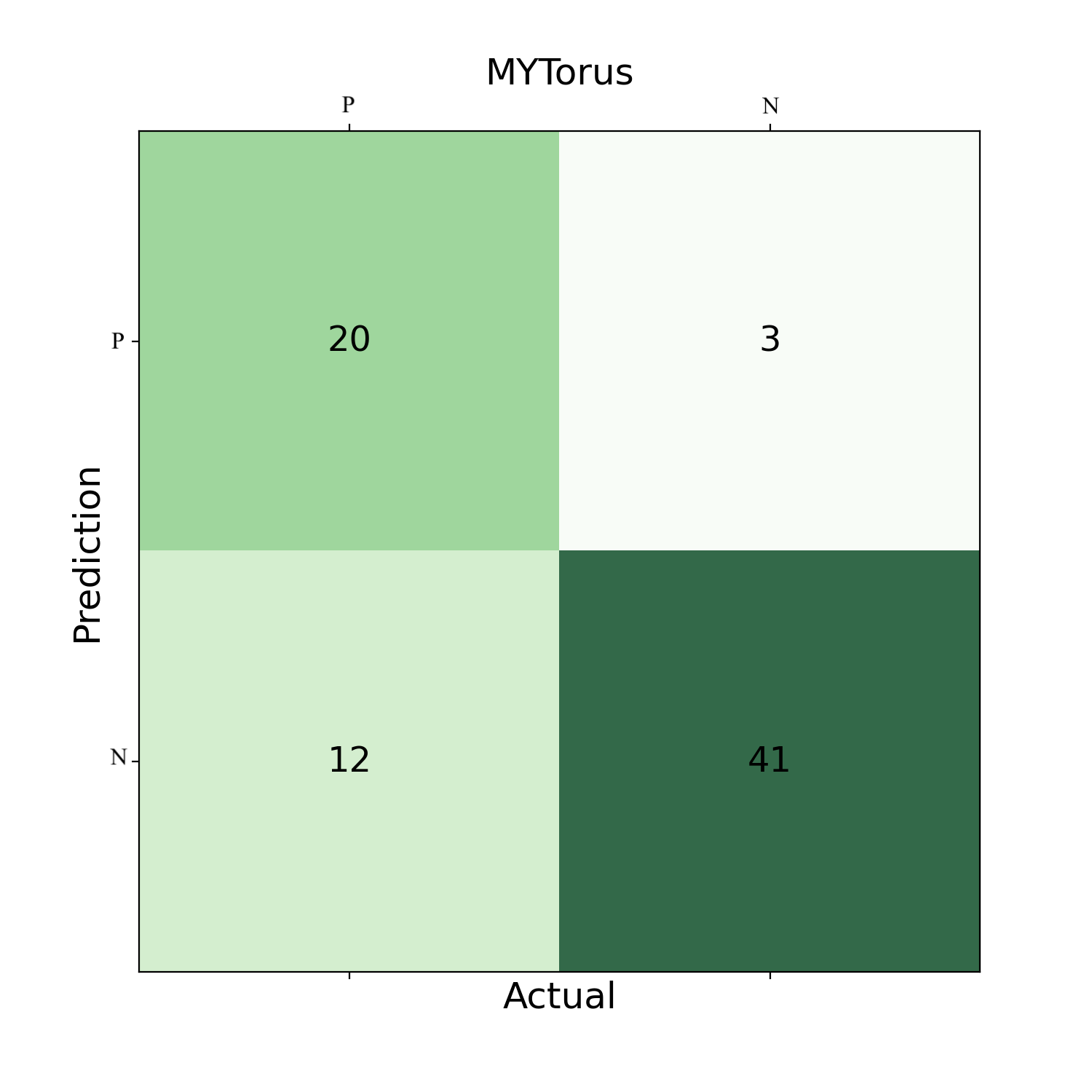}\hfill
    \includegraphics[width=0.33\textwidth]{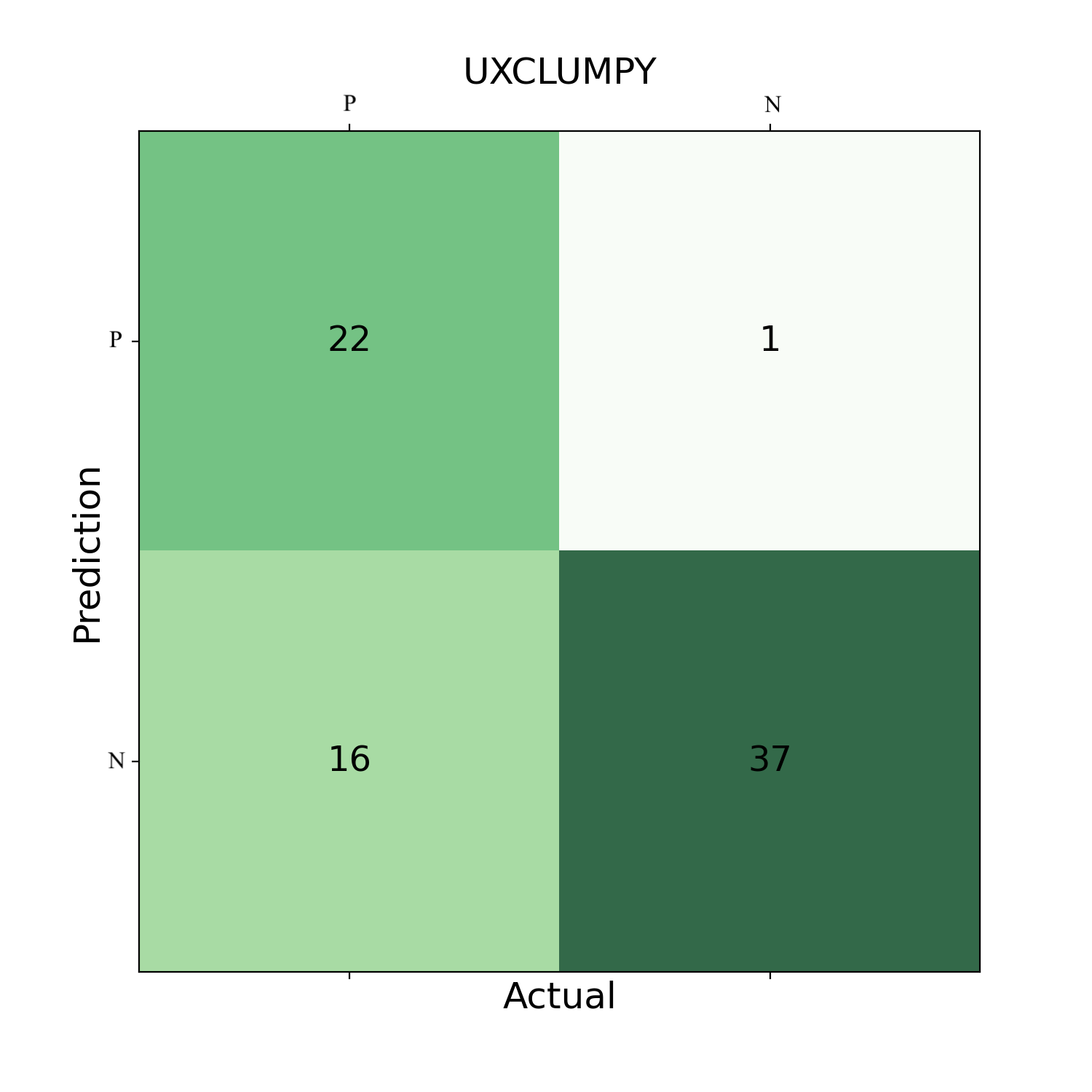}\hfill
    \caption{Confusion matrices for all three models using Criterion\,2 at a 99.999\,\% confidence level considering $\textit{HR}_1$\&$\textit{HR}_2$. These show that the method is good at avoiding false positives meaning that almost none of the non-variable sources in a sample will be flagged. However, this comes at the expense of not selecting a larger number of variable sources (bottom left).}
    \label{fig:B3}
\end{figure*}

Figures \ref{fig:B1}$-$\ref{fig:B3} show the confusion matrices for all three models and both criteria considering both $\textit{HR}_1$\&$\textit{HR}_2$. The diagonal of these matrices are the correct predictions while the top right are false positives (non-variable sources classified as variable) and the the bottom left are false negatives (variable sources classified as non-variable). Figure \ref{fig:B2} shows the results considering a 90\,\% confidence level in Criterion\,2. This corresponds to the red dots in Figure \ref{fig:roc}. Figure \ref{fig:B3} shows the results considering a 99.999\,\% confidence level in Criterion\,2. This corresponds to the far left side of the red lines in Figure \ref{fig:roc}.

\end{document}